\documentclass[journal]{IEEEtran}
\usepackage{graphicx}
\usepackage{comment}
\usepackage{amsmath,amssymb} 
\usepackage{color}
\graphicspath{{figs1/}}
\usepackage{multirow}
\usepackage{caption}
\usepackage{hyperref}

\ifCLASSINFOpdf
\else
\fi

\makeatletter
\newcommand{\printfnsymbol}[1]{%
	\textsuperscript{\@fnsymbol{#1}}%
}
\makeatother
\begin{document}

\title{Exploring Overcomplete Representations for Single Image Deraining using CNNs}

\author{
		Rajeev Yasarla\printfnsymbol{1}\thanks{*equal contribution},~\IEEEmembership{Student Member,~IEEE,} Jeya Maria Jose Valanarasu\printfnsymbol{1},~\IEEEmembership{Student Member,~IEEE,} and Vishal~M.~Patel,~\IEEEmembership{Senior Member~,~IEEE}
		\thanks{Rajeev Yasarla, is with the Whiting School of Engineering, Johns Hopkins University, 3400 North Charles Street, Baltimore, MD 21218-2608, e-mail: ryasarl1@jhu.edu}
		\thanks{Jeya Maria Jose Valanarasu, is with the Whiting School of Engineering, Johns Hopkins University, 3400 North Charles Street, Baltimore, MD 21218-2608, e-mail: jvalana1@jhu.edu
			\footnotesize}
		
		\thanks{Vishal M. Patel, is with the Whiting School of Engineering, Johns Hopkins University, e-mail: vpatel36@jhu.edu}
}

\maketitle

\begin{abstract}
	Removal of rain streaks from a single image is an extremely challenging problem since the rainy images often contain rain streaks of different size, shape, direction and density.  Most recent methods for deraining use a deep network following a generic ``encoder-decoder" architecture which captures low-level features across the initial layers and high-level features in the deeper layers. For the task of deraining, the rain streaks which are to be removed are relatively small and focusing much on global features is not an efficient way to solve the problem. To this end, we propose using an overcomplete convolutional network architecture which gives special attention in learning local structures by restraining the receptive field of filters. We combine it with U-Net so that it does not lose out on the global structures as well while focusing more on  low-level features, to compute the derained image. The proposed network called, Over-and-Under Complete Deraining Network (OUCD), consists of two branches: overcomplete branch which is confined to small receptive field size in order to focus on the local structures and an undercomplete branch that has larger receptive fields to primarily focus on global structures.  Extensive experiments on synthetic and real datasets demonstrate that the proposed method achieves significant improvements over the recent state-of-the-art methods. Code is available at \url{https://github.com/jeya-maria-jose/Derain_OUCD_Net}
\end{abstract}


%
\IEEEpeerreviewmaketitle

\section{Introduction}

Images taken in outdoors are more susceptible to different weather conditions which decrease the visual quality of the captured images. This results in degradation of many high-level computer vision tasks like video surveillance, image understanding, object detection and classification. In order to improve the performance of these tasks, it is important to develop algorithms that remove artifacts and recover the cleaner version of the degraded images. Rain is one of the most common weather condition that is responsible for poor performance of high-level computer vision algorithms in video surveillance and autonomous driving \cite{yang2019single}.  Removing rain streaks from the rainy images can considerably increase the performance of these vision tasks \cite{zhang2019image}. 

Deep Learning methods have risen to popularity in solving most of the  computer vision tasks in the last decade. They have achieved state-of-the-art performance in many tasks like image classification, object detection, semantic segmentation and image restoration. Most of these methods are based on convolutional neural networks (CNNs). AlexNet \cite{krizhevsky2012imagenet} in 2012 showed how a deep network consisting of convolutional,  max-pooling and fully-connected layers works very well for image classification task. Most of the methods that followed \cite{szegedy2015going,he2016deep} used a similar set of layers with changes in architecture for efficient learning. Seg-Net \cite{badrinarayanan2017segnet}, an ``encoder-decoder" type of deep network was proposed for the task of semantic segmentation which gave a significant improvement in the performance over previous methods. The main idea behind an ``encoder-decoder" convolutional network architecture is that at the initial convolutional layers, the receptive field size of the filters are small and so capture low-level features like edges. The deeper layers in the network work on larger receptive field and so capture high-level features of the input image. The feature maps are projected onto a lower dimension similar to an undercomplete autoencoder. The decoder takes the feature maps back to the original resolution of the input image. U-Net \cite{ronneberger2015u} proposed using skip connections between the encoder and decoder network which showed a significant improvement in performance for image segmentation. Following the popularity of U-Net, it was the go-to network architecture serving as the backbone for many other tasks like image-to-image translation \cite{isola2017image,yi2017dualgan,zhang2017style}, image synthesis \cite{wang2018high}, image segmentation \cite{iglovikov2018ternausnet}, medical image analysis \cite{jin2017deep,milletari2016v,zhou2018unet++,cciccek20163d} and image restoration \cite{yang2019single,yasarla2020deblurring,yasarla2020learning}. 

\begin{figure*}[htp!]
	\begin{center}
		\centering
		\includegraphics[width=2.7cm,height=2cm]{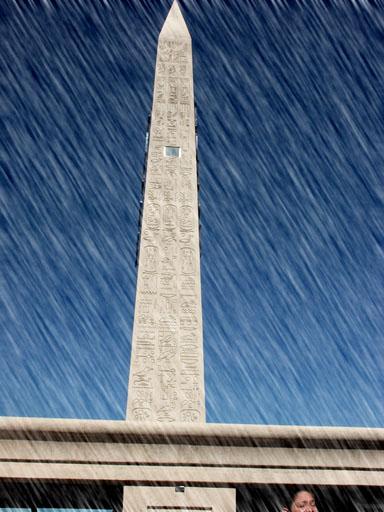}
		\includegraphics[width=2.7cm,height=2cm]{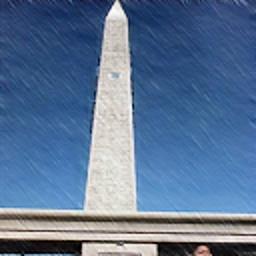}
		\includegraphics[width=2.7cm,height=2cm]{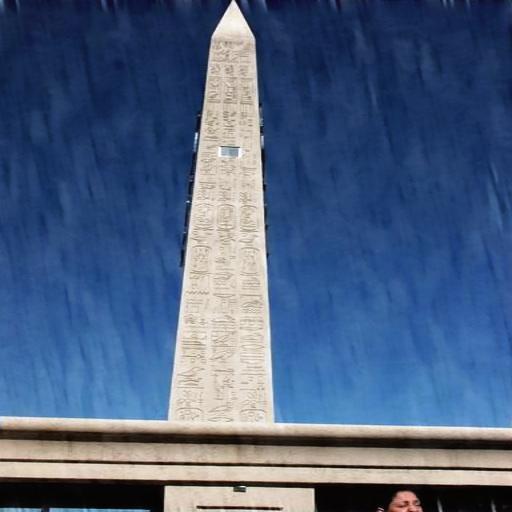}
		\includegraphics[width=2.7cm,height=2cm]{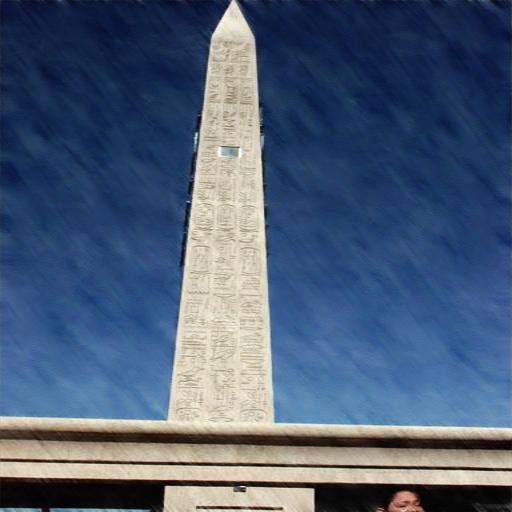}\\
		Rainy Image \hskip30pt UNet~\cite{ronneberger2015u} \hskip 45pt DDN~\cite{fu2017removing} \hskip35pt DIDMDN~\cite{zhang2018density}\\
		\includegraphics[width=2.7cm,height=2cm]{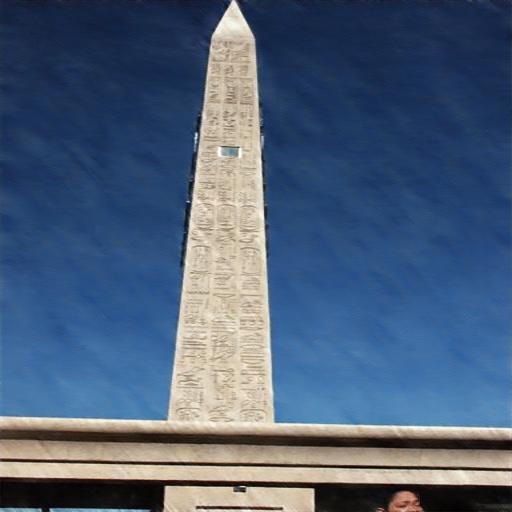}
		\includegraphics[width=2.7cm,height=2cm]{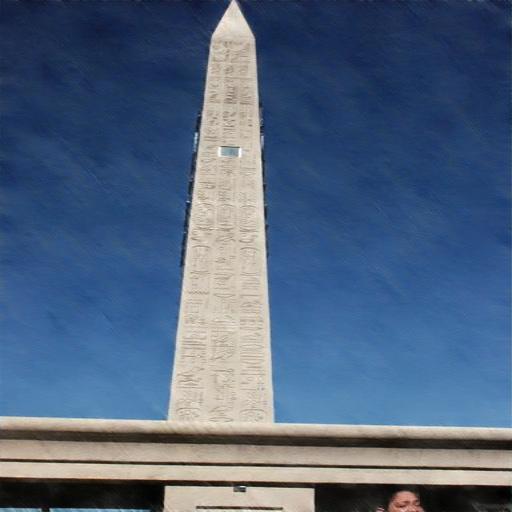}
		\includegraphics[width=2.7cm,height=2cm]{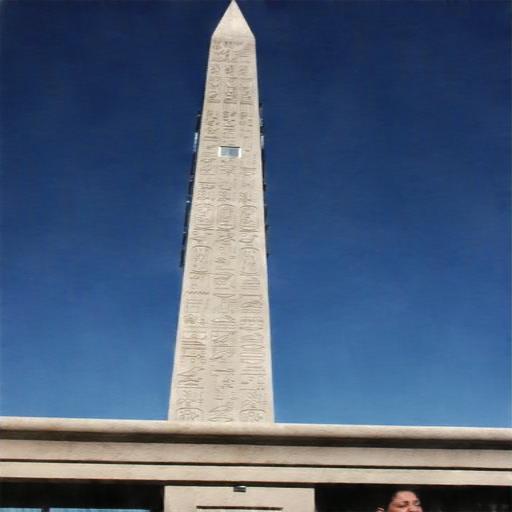}
		\includegraphics[width=2.7cm,height=2cm]{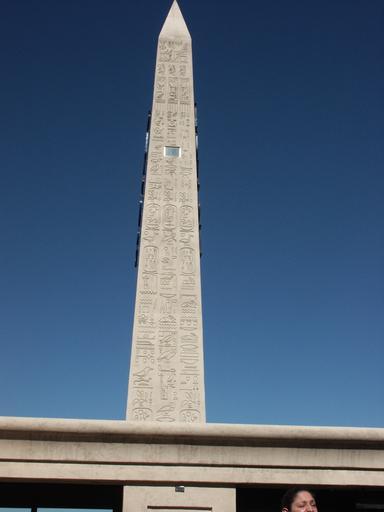}\\
		RESCAN~\cite{li2018recurrent} \hskip30pt SPAnet~\cite{Wang_2019_CVPR} \hskip 30pt OUCD (ours) \hskip25pt Clean Image\\
		\includegraphics[width=2.7cm,height=2cm]{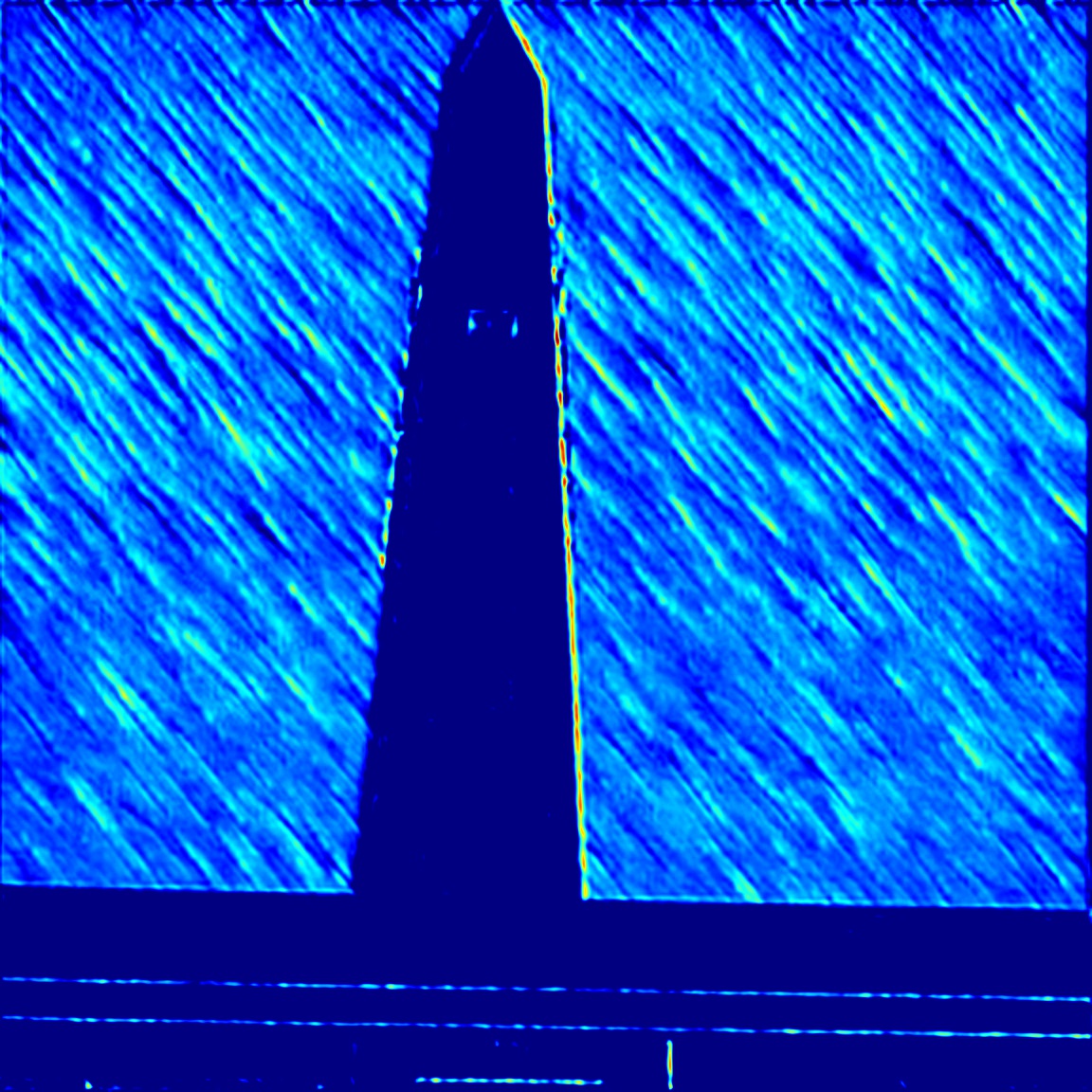}
		\includegraphics[width=2.7cm,height=2cm]{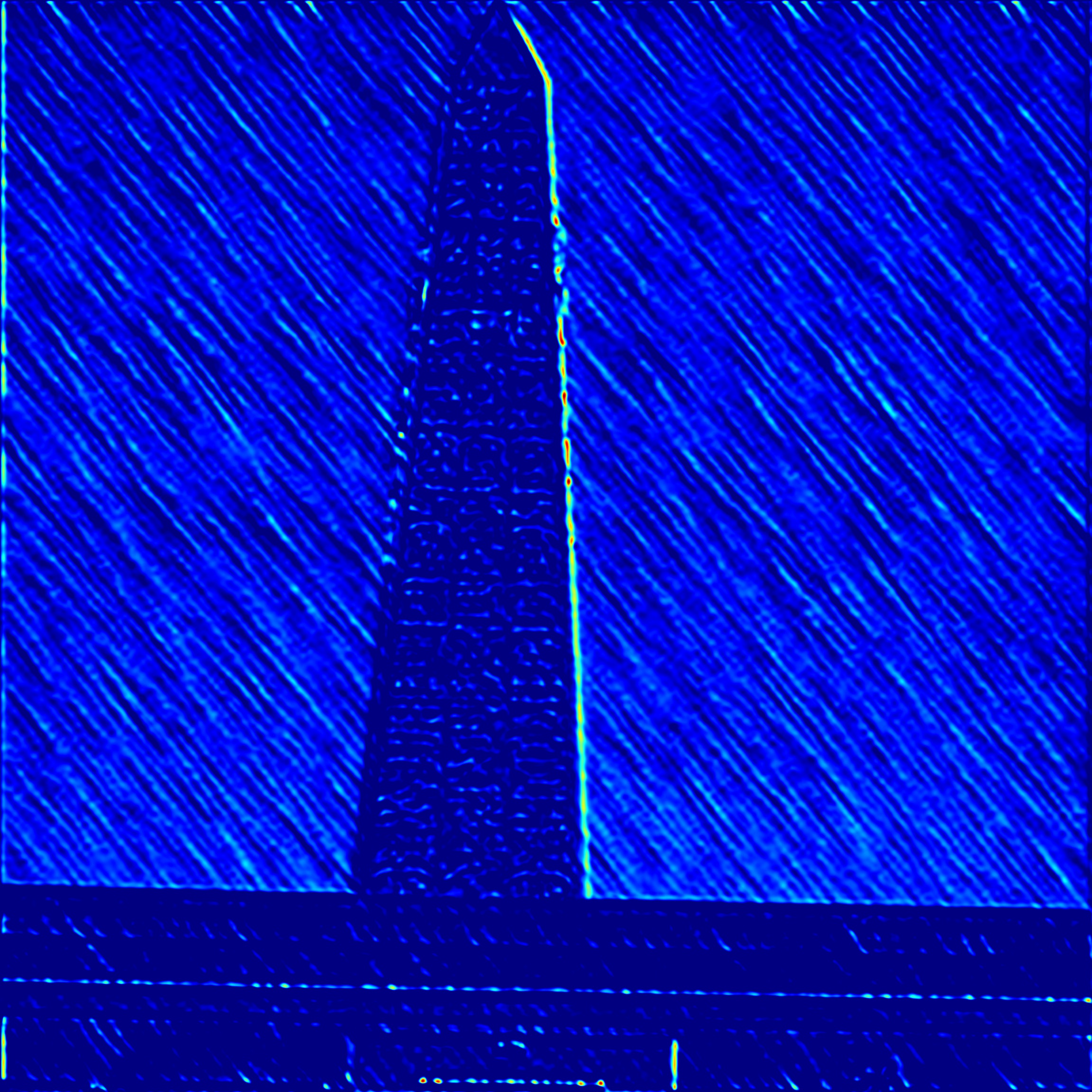}
		\includegraphics[width=2.7cm,height=2cm]{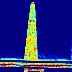}
		\includegraphics[width=2.7cm,height=2cm]{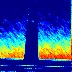}\\
		(a) \hskip65pt (b) \hskip70pt (c) \hskip65pt (d)\\
		\caption{First and Second rows show the comparison of our method OUCD, for the rainy image against the state-of-the-art methods. It can be seen that our proposed method OUCD produces a better derained image when compared to the other methods. Third row: (a),(b) Features maps from the deep layers of an overcomplete network. (c),(d) Feature maps from the deep layers of an undercomplete network. The feature maps of an overcomplete network captures rain streaks as seen in (a),(b) while the undercomplete network captures details of the building or the sky as seen in (c),(d). }
		\label{Fig:quali_exp1}
	\end{center}
	\vskip-15pt
\end{figure*}

The noisy element can be different in each restoration problem, it can be small streaks in case of deraining, it can be larger patches in case of shadow removal or it can be grains of speckles in case of denoising problem. Using a generic ``encoder-decoder" architecture like U-Net or its variant works well in most of the restoration problems as it captures both low-level features (in initial layers) and high-level features (in deeper layers). However, there is a difference in the features that are to be given importance according to the restoration task at hand. In tasks like deraining where most of the rain streaks in the image are small; using a filter that has a very large receptive field is not useful as it does not learn any information about the rain streaks. For example, Fig.~\ref{Fig:quali_exp1} shows that existing deraining methods based on ResNet and U-Net architectures are not able to remove rain streaks completely since they are not able to focus on meaningful low-level features. \cite{fu2017removing,li2018recurrent,Wang_2019_CVPR} which are based on ResNet architecture, fail to remove small rain streaks and fail to capture rain streaks in the sky as seen in Fig. \ref{Fig:quali_exp1}.  \cite{ronneberger2015u,zhang2018density} which are U-Net based architectures primarily focus on global level features and fail to remove small rain streaks and also  remove some texture data as explained in \cite{Wang_2019_CVPR,yasarla2019uncertainty}.  In this paper, we focus on extracting meaningful low-level features that can capture even the tiniest rain streaks and remove them. To this end, we propose to use an overcomplete convolutional network architecture. 

Before the deep learning era, overcomplete representations  \cite{lewicki2000learning} have been used for many tasks where a robust method that could deal with noisy input was needed. Although overcomplete architectures have been mentioned in the literature \cite{vincent2008extracting}, overcomplete convolutional architectures have not been explored for low-level vision tasks. Unlike an undercomplete convolutional network where the receptive field enlarges as we go deeper through the network, an overcomplete architecture restricts the enlargement of receptive field in the deeper layers (the reason is discussed in section~\ref{focus}). This helps us to force the filters in all the layers of network focus on low-level features. Moreover, the deeper layers now learn more finer edges and focus on smaller objects than the initial layers which helps us to detect and eliminate rain streaks better. Though we state that low-level features are more important for the task of deraining, it does not mean that high-level features do not carry any meaning at all for this task. High level-features are also important as they help in properly reconstructing large structures and objects in the image. Hence, we propose a novel network architecture that combines both the overcomplete and undercomplete architectures in an efficient way for image deraining. We name our network Over-and-Under Complete Deraining (OUCD) Network.  The overcomplete branch of our network focuses on learning low level features in high resolution thus capturing very fine features of rain streaks, while the undercomplete branch learns global features. Some of the feature maps from our network are illustrated in Fig. \ref{Fig:quali_exp1} (a),(b),(c) and (d). From these maps, it can be clearly seen that an overcomplete convolutional network branch captures the rain streaks clearly even in their deep layers as seen in Fig. \ref{Fig:quali_exp1} (a),(b) whereas the undercomplete network branch captures the building and sky in its deep layers as can be seen in Fig.~\ref{Fig:quali_exp1} (c) and (d).   

To summarize, this paper makes the following contributions:
\begin{itemize}
	\item We explore overcomplete representations in deep learning network architectures for image deraining.
	\item We show how using an overcomplete architecture helps in capturing finer details of rain streaks more effectively.
	\item We propose a novel deep network architecture,  Over-and-Under Complete Deraining (OUCD) Network, which combines both overcomplete and undercomplete networks so as to learn both low-level and high-level features of the rainy image effectively.
	\item We achieve significant improvements in performance across many real and synthetic deraining datasets over the recent state-of-the-art methods.
\end{itemize}

\section{Related Work}\label{relatedworks}
We have witnessed great progress in the literature addressing the rain removal problems  using the traditional image processing methods \cite{kang2011automatic,chen2014visual,luo2015removing,li2016rain} and convolutional neural networks (CNNs) \cite{zhang2017convolutional,wang2017hierarchical}. These rain removal methods can be broadly divided into video rain removal methods \cite{zhang2006rain,garg2007vision,santhaseelan2015utilizing,liu2018erase,li2018video,yang2019frame}, and single-image rain removal methods \cite{kang2011automatic,li2016rain,ren2019progressive,wang2019erl,Wang_2019_CVPR,deng2020detail,wang2020model,jiang2020multi,yasarla2020confidence,du2020conditional,wang2020rethinking}. In this paper, we focus on single image rain removal problem. A captured rainy image $y$ is mathematically modeled as the addition of rain streak information $r$ (i.e called as residual map) with the clean image $x$ as follows
\begin{equation}
y = x + r.
\end{equation}

\noindent {\bf{Video-based methods:}} Video-based deraining methods are rich in temporal information, so one can use temporal consistency to ease the deraining problem. \cite{zhang2006rain,garg2007vision,santhaseelan2015utilizing} exploit temporal information in different ways and make use of them to remove the rain information in videos. \cite{li2018video} proposed a method using multi-scale convolutional sparse coding model for the video rain streak removal that takes both characteristics of rain information into account like rain streaks follows multi-scale configurations and their patterns are sparsely scattered. \cite{yang2019frame} proposed a two-step architecture where they extract reliable motion information from the initially estimated derained image to align the frames and then model the motion in the second stage.\\

\noindent {\bf{Single-image based methods:}} Early  methods in the literature like \cite{kang2011automatic,chen2014visual,luo2015removing,li2016rain} for the rain removal task use traditional image processing techniques. They include techniques that address the rain removal problem using low-high frequency image decomposition~\cite{kang2011automatic}, low-rank representation based~\cite{liu2012robust},  dictionary learning-based~\cite{mairal2009online}, and Gaussian mixture model-based methods~\cite{li2016rain}. Fu \textit{et al.} \cite{fu2017removing,fu2017clearing} introduced CNN-based methods in an end-to-end deep learning fashion to address the rain removal problem. Zhang \textit{et al.}~\cite{zhang2018density} proposed density aware multi-stream CNN where first they estimate density of rainy image and use it to fuse the features from different streams in order to derain the image. Qian \textit{et al.} proposed an attention based rain drop removal network where they use recurrent network based architecture to obtain the attention map. Li \textit{et al.}~\cite{li2018recurrent} exploits the benefits of recurrent neural networks to preserve the useful contextual information of the earlier layers. Ren \textit{et al.}~\cite{ren2019progressive} proposed a simpler baseline network that is based on ResNet and performs deraining in a progressive manner to obtain the final derained image. \cite{hu2019depth,halder2019physics} proposed rainy datasets that considers physics of rainy images formulation. Wang \textit{et al.}\cite{Wang_2019_CVPR} proposed a local to global procedure to derain the image using spatially attentive blocks. Yasarla \textit{et al.} \cite{yasarla2019uncertainty} proposed a  mutliscale  deraining method by modeling uncertainity in estimating residual at low scales and using them in computing the final derained image. Wei \textit{et al.}~\cite{wei2019semi} proposed a semi-supervised deraining method using Gaussian mixture models. Wang  \textit{et al.}\cite{wang2019erl} proposed a network that learns a mapping from low-quality embedding to a latent optimal vector by formulating a residual learning branch, that is capable of adaptively adding residuals between the original low-quality embedding  to the latent vector in a entanglement representation manner. Yasarla \textit{et al.}  \cite{yasarla2020syn2real} proposed a semi-supervised method to leverage rain information from real rain images in training the network using Guassian processes. Very recent works that propose using techniques like pyramid networks, contextual deep networks for image deraining can be found in \cite{8767931,WangA,8627954}.

\section{Proposed Method}\label{focus}

As discussed earlier, most image deraining methods are based on either an ``encoder-decoder" architecture (U-Net based) or a residual architecture.  These architectures do not focus much on the local features as the deeper layers of these networks have large receptive fields thus extracting high-level features. Although learning global-features is one of the main reasons an encoder-decoder architecture was used in the first place, it makes sense to use it for tasks like object detection, classification and segmentation as the object of interest in the image is comparatively larger. But in the task of deraining, the rain streaks are mostly small and are better detected when we use filters of small receptive field. To this end, we force our network to learn low-level features by restricting the receptive field from enlarging in the deeper layers of the network by using an overcomplete convolutional architecture. In an overcomplete architecture, the input image is taken into a higher dimension (spatially) unlike the undercomplete architecture. This happens as the max-pooling layers in an undercomplete architecture are replaced by upsampling layers. The receptive field increases in an undercomplete architecture by using a max-pooling layer along with a convolutional layer. During feed-forward, the feature maps of size $k \times k$ that go through a pooling layer of a coefficient $n$ become $(k/n) \times (k/n)$. However, during back propagation, the flow of gradients go from a kernel of size $k \times k$ to $nk \times nk$. This increases the receptive field size of the filters in the deeper layers of an undercomplete network resulting in the features to learn high-level global features of the image. In an overcomplete architecture, if we use an upsampling layer instead of max-pooling layer, we get an inverse effect of what happened before. The feature maps go from a size of $k \times k$ to the size of $nk \times nk$ after every upsampling layer. Thus, during back-propagation, the flow of gradients go from $k \times k$ to $(k/n)\times(k/n)$. Although each convolutional layer causes the receptive field to enlarge (depending on the kernel size of convolutional layer), the upsampling layer subdues this effect to some extent. Overall, an overcomplete convolutional architecture forces the receptive field to not increase as much as an undercomplete architecure does; resulting in the filters across all layers to learn low-level features.

\begin{figure*}
	\centering
	\includegraphics[width=0.8\linewidth]{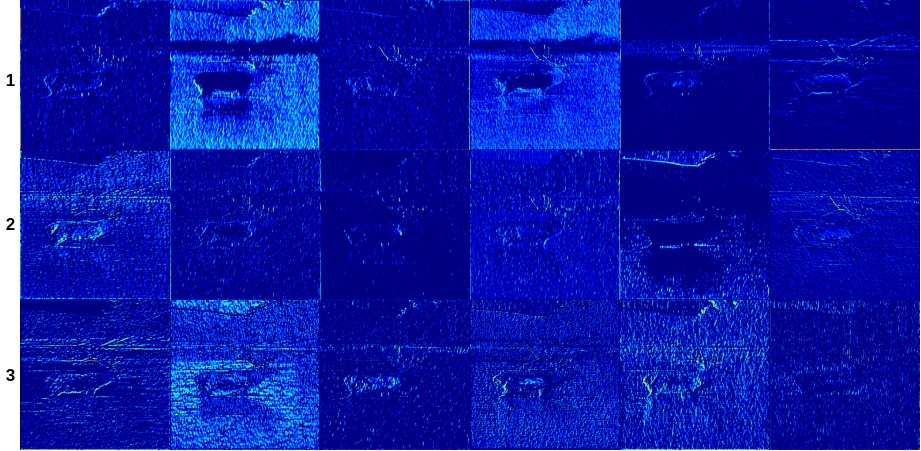}
	\caption{Feature maps in an overcomplete architecture. The rows correspond to the encoder layer from which the feature maps are taken. It can be seen that the overcomplete network captures information in the image which are mainly local and so we can see a lot of filters here capture fine details like rain streaks and drops. }
	\label{onetfilters-fig}
\end{figure*}

\begin{figure*}
	\centering
	\includegraphics[width=0.8\linewidth]{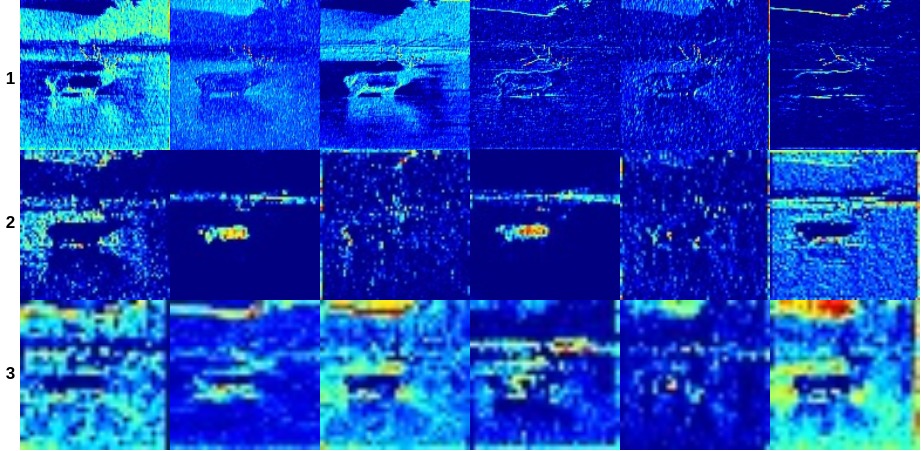}
	\caption{Feature maps in an undercomplete architecture. The rows correspond to the encoder layer from which the feature maps are taken. The undercomplete network captures information in the image which high-level and so we can see that a lot of filters here capture details like the deer and its background. }
	\label{unetfilters-fig}
\end{figure*}

In Fig. \ref{onetfilters-fig}, we illustrate the feature maps of the overcomplete network taken across different layers when it is trained for deraining. We can observe that all the layers of the network learn to capture low-level features of the rain streaks properly. It can also be noted here that as we go deeper in the overcomplete architecture, the features captured are more finer as the receptive field is small and acts at a higher resolution of the input. In Fig. \ref{unetfilters-fig}, we illustrate the feature maps across different layers of an undercomplete U-Net architecture. It can be seen that the feature maps in the deeper layers focus on learning global features which are not necessarily rain streaks. So, it is evident that for tasks such as deraining, low level features are much more useful and so we give more importance to low-level feature extraction in our proposed method as explained in the next sub-section.

\subsection{OUCD Network}
\begin{figure*}
	\centering
	\includegraphics[width=0.8\linewidth]{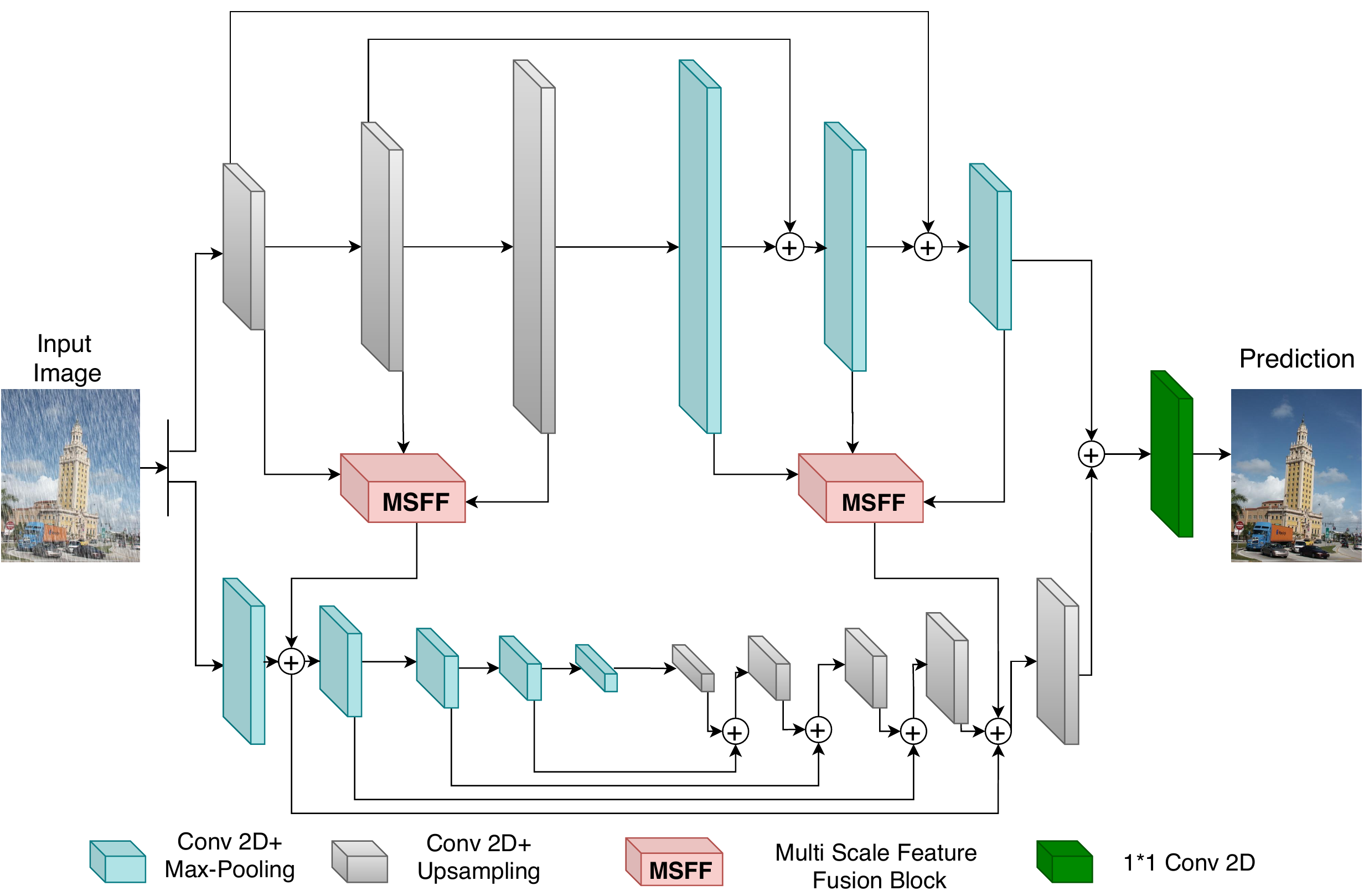}
	\caption{An overview of the proposed OUCD network architecture.}
	\label{derainarch}
\end{figure*}

Fig.~\ref{derainarch} gives an overview of the proposed OUCD network.  We use two  network branches: an overcomplete and an undercomplete branch. The intuition behind using both the architectures is to make use of both local and global features. Although we focus more on the local features using the overcomplete branch, we do not leave out the global features altogether as they still have some meaningful information helpful for proper restoration of the image. In the overcomplete branch, we have 3 convolutional blocks in both encoder and decoder. Each convolutional block in the encoder has a 2D convolutional layer followed by an upsampling layer and ReLU activation. All the convolutional layers in our architecture have a kernel size of $3\times 3$, stride of 1 and padding of 1 unless mentioned otherwise. For upsampling, we perform bilinear upsampling with a scale factor of 2. In the decoder, each convolutional block has a 2D convolutional layer followed by max-pooling layer and ReLU activation \cite{nair2010rectified}. The max-pooling layer has a pooling coefficient of 2. We also have skip connections similar to U-Net \cite{ronneberger2015u} from the encoder layers to the decoder layers for better localization. 

The undercomplete branch is similar to the standard U-Net architecture. It has 5 convolutional blocks in both encoder and decoder. Each convolutional block in the encoder consists of a 2D convolutional layer followed by a max-pooling layer and ReLU activation. Each convolutional block in the decoder consists of a 2D convolutional layer followed by an upsampling layer and ReLU activation. Both pooling and upsampling layers have coefficients of  two. We also have skip connections from each block of the encoder to the corresponding block in the decoder, similar to U-Net architecture. More details regarding the network architectures corresponding to the overcomplete and undercomplete branches are provided in the Appendix.

As we have established that overcomplete branch extracts useful low-level features across all its layers, we propose to use all those feature maps for a better prediction for the deraining task as most of the rain streaks are low-level features. These feature maps capture very fine details than the initial layers of an undercomplete branch (U-Net) as the resolution of the image that the overcomplete network acts on is very high when compared to the initial layer of U-Net. So, we make use of all of these feature maps from the encoder of the overcomplete branch by adding them to the output of the first layer of the undercomplete branch. This helps in the later layers of the undercomplete branch to learn better global features. Similarly, all the feature maps in the decoder of overcomplete branch are added to the features maps of the undercomplete branch's decoder just before the last block. This facilitates better restoration at the decoder part as well. Before adding the feature maps from overcomplete branch, we pass it through Multi-Scale-Feature-Fusion (MSFF) block. The details of the MSFF block are given in the next subsection. Moreover, we add the last set of feature maps from the overcomplete branch with undercomplete branch before passing it on to the last convolutional layer. The last convolutional layer has $1\times 1$ kernels to convert the feature maps into a 3-channel RGB image. This prediction is then compared with ground-truth using a loss function to calculate the gradients for back-propagation.

\subsection{MSFF Block}
\begin{figure}[htp!]
	\centering
	\includegraphics[width=\linewidth]{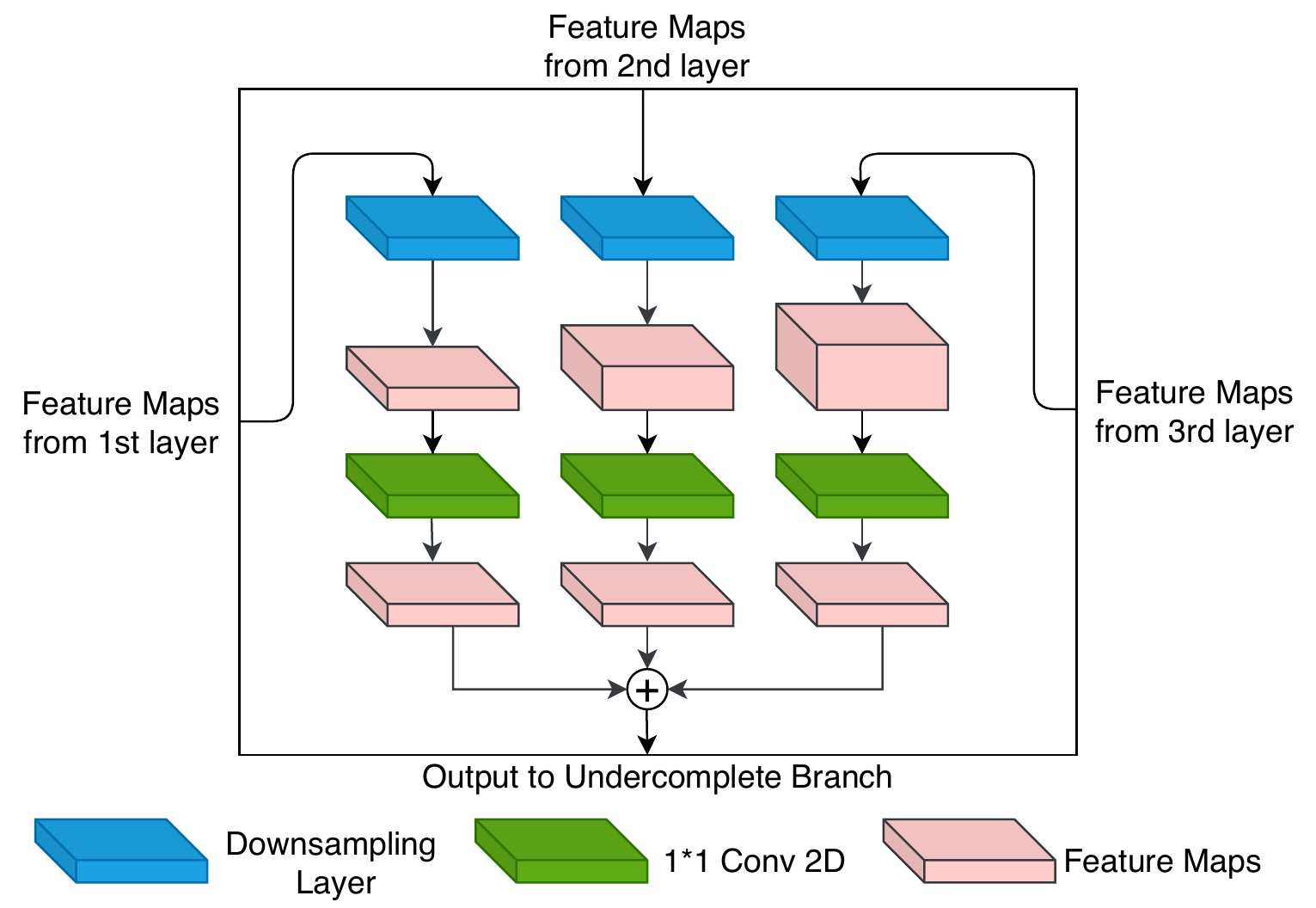}
	\caption{The architecture of Multi-Scale-Feature-Fusion (MSFF) Block.}
	\label{msff}
\end{figure}

We propose an MSFF block to transfer feature maps from different scales of the overcomplete branch to the undercomplete branch,  transform them into the similar scales while also maintaining  equal number of feature maps such that they have equal weightage when they are added.  Fig. \ref{msff} illustrates the network architecture of the MSFF block.  We use a convolutional block that consists of a downsampling layer followed by a $1\times 1$ convolutional layer across each scale. The downsampling layer is used to downsample the feature maps into a scale as that of feature maps in the undercomplete branch it is going to be added to. We use bilinear interpolation for the downsampling operation. The scale factor for each scale is equal to the ratio of the size of feature map in the undercomplete branch to the size of the feature map at that particular scale. The $1\times 1$ convolution layer is used to make the number of feature maps taken across different scales of the overcomplete network to be consistent so that all the feature maps have equal weightage when they are getting added to the undercomplete branch. Then the feature maps are added and are passed on to the undercomplete branch.\\

\noindent {\bf{Loss Function:}} 
We use the standard $ \ell_{2} $ loss calculated between the predicted $\hat{x}$ and the ground truth $x$.  It is defined as follows 

\[\mathcal{L}_{mse} = \lVert \hat{x}-x \rVert _{2}^{2}. \] We also compute a perceptual loss \cite{johnson2016perceptual} using a pretrained VGG-16 network. It is defined as follows \[\mathcal{L}_{p} = \frac{1}{NHW}\sum_i \sum_j \sum_k \lVert F(\hat{x})^{i,j,k} - F(x)^{i,j,k} \rVert _{2}^{2}, \]
where $N$ is the number of channels of $F(.)$, and $H$ and $W$ denote the height and the width of the feature maps.  Here, $x$ denotes the ground truth and $\hat{x}$ denotes the predicted derained image.  We take features from $relu1\_2$, $relu2\_2$ and $relu3\_2$ of VGG-16 for computing the perceptual loss.
The overall loss used to train the OUCD network is,
\begin{equation}\label{eq:loss} \mathcal{L} = \mathcal{L}_{mse} + \lambda \mathcal{L}_{p}, \end{equation}
where $\lambda$ is a parameter which we set equal to 0.04 in our experiments.

\section{Experiments and Results}

In this section, we conduct experiments to compare the proposed OUCD method  with state-of-the-art methods quantatively and qualitatively on both synthetic rain and real-world rainy images. We also conduct extensive ablation study to show the benefits of our proposed method over the U-Net and overcomplete base networks. We use Peak-Signal-to-Noise Ratio (PSNR) and Structural Similarity index (SSIM) as the comparative measures to evaluate the performance of different methods. For qualitative comparision and analysis, we visually present the sample results of different methods. We compare OUCD against the following state-of-the-art methods,
\begin{itemize}
	\item Discriminative sparse coding (DSC) \cite{luo2015removing}(ICCV'15)
	\item Gaussian mixture model (GMM) (GMM) \cite{li2016rain} (CVPR'16)
	\item Joint Rain Detection and Removal~(JORDER) \cite{yang2017deep} (CVPR'17)
	\item Deep detailed Network (DDN)\cite{fu2017removing}(CVPR'17)
	\item REcurrent SE Context Aggregation Net (RESCAN) \cite{li2018recurrent} (ECCV'18)
	\item SPatial Attentive Network (SPANet)\cite{Wang_2019_CVPR}(CVPR'19)
	\item PreNet \cite{ren2019progressive} (CVPR'19).\\
\end{itemize}

\noindent {\bf{Training Dataset:}} 
We train the proposed network using synthetic rain datasets \cite{zhang2018density,zhang2019image} and real rain dataset \cite{Wang_2019_CVPR}. The DIDMDN synthetic rain dataset published by the authors of \cite{zhang2018density}, contains 4000 low density rain, 4000  medium density rain, and 4000 high density rain images for training. The Rain800 \cite{zhang2019image} synthetic rain dataset contains 700 training images. We also train our network on the real rain dataset proposed by the authors of SPANet \cite{Wang_2019_CVPR} which contains 342 real rain frames for training.\\

\noindent {\bf{Test Dataset:}} We evaluate the performance of OUCD against the state-of-the-art methods on the synthetic and real rain test datasets published by the authors of \cite{li2016rain,zhang2018density,Wang_2019_CVPR,zhang2019image}. DIDMDN \cite{zhang2018density} contains two test sets.  Test-1 has a total of 1200 images where it is a combination of 400 low density, 400 medium density, and 400 high density images. Test-2 has a total 1000 images where the images are randomly sampled from the synthetic test set of Fu \textit{et al.}\cite{fu2017removing}.  The Rain12 dataset contains 12 images shared by the authors of \cite{li2016rain}. The Rain800 test set provided by the authors of \cite{zhang2019image} contains 101 synthetic rain images. Authors of  SPANet\cite{Wang_2019_CVPR} published a real-world test set which contains 1000 real rainy images for testing. In addition to these test datasets,  we also perform qualitative comparsions of different methods using real-world rainy images provided by the authors of \cite{zhang2019image,Wang_2019_CVPR}.\\

\noindent {\bf{Implementation Details:}}
We train the OUCD network with pairs of rainy and the corresponding clean images $\{y,x\}$,  using the loss $\mathcal{L}$ defined in \eqref{eq:loss}. We augment the input rainy image by randomly cropping $128\times 128$ patches for training the network. We use the Adam optimizer \cite{kingma2014adam} with the batchsize of 2 for optimizing the loss. We set the learning rate equal to 0.0002 for 30 epochs and 0.0001 for the remaining epochs. We implemented our network on the PyTorch framework and perform all the experiments using two NVIDIA RTX 2080Ti GPUs.

\begin{table*}[htp!]
	\centering
	\caption{PSNR and SSIM (PSNR$|$SSIM) quantitative comparison of OUCD against the state-of-the-art methods.}
	\label{experiments}
	\resizebox{\textwidth}{!}{
		\begin{tabular}{|l|l|c|c|c|c|c|c|c|c|c|}
			\hline
			\multicolumn{2}{|l|}{Datasets} & \begin{tabular}[c]{@{}c@{}}DSC\cite{luo2015removing}\\ (ICCV'15)\end{tabular} & \begin{tabular}[c]{@{}c@{}}GMM\cite{li2016rain}\\ (CVPR'16)\end{tabular} & \begin{tabular}[c]{@{}c@{}}JORDER\cite{yang2017deep}\\ (CVPR'17)\end{tabular} & \begin{tabular}[c]{@{}c@{}}DDN\cite{fu2017removing}\\ (CVPR'17)\end{tabular} & \begin{tabular}[c]{@{}c@{}}DIDMDN\cite{zhang2018density}\\ (CVPR'18)\end{tabular} & \begin{tabular}[c]{@{}c@{}}RESCAN\cite{li2018recurrent}\\ (ECCV'18)\end{tabular} & \begin{tabular}[c]{@{}c@{}}SPANet\cite{Wang_2019_CVPR}\\ (CVPR'19)\end{tabular} & \begin{tabular}[c]{@{}c@{}}PreNet\cite{ren2019progressive}\\ (CVPR'19)\end{tabular} & \begin{tabular}[c]{@{}c@{}}OUCD\\ (ours)\end{tabular} \\ \hline
			\multirow{2}{*}{DIDMDN} & Test-1 & 21.44$|$0.79 & 22.75$|$0.84 & 24.32$|$0.86 & 27.33$|$0.90 & 27.95$|$0.91 & 27.19$|$0.87 & 30.05$|$0.91 & 31.26$|$0.91& \textbf{31.49}$|$\textbf{0.91} \\ \cline{2-11} 
			& Test-2 & 20.08$|$0.78 & 20.66$|$0.81 & 22.26$|$0.84 & 25.63$|$0.89 & 26.07$|$0.91 & 25.65$|$0.88 & 26.19$|$0.90 & 26.27$|$0.90& \textbf{26.73}$|$\textbf{0.92} \\ \hline
			\multicolumn{2}{|l|}{SPANet} & 32.33$|$0.93 & 32.99$|$0.95 & 35.72$|$0.97 & 33.28$|$0.97 & 28.96$|$0.95 & 35.19$|$0.98 & 38.06$|$0.98 & 38.55$|$0.98 & \textbf{39.25}$|$\textbf{0.98} \\ \hline
			\multicolumn{2}{|l|}{Rain12} & 30.07$|$0.85 & 32.02$|$0.86 & 33.92$|$0.95 & 31.78$|$0.90 & 31.32$|$0.90 & 32.42$|$0.92 & 34.12$|$0.94 & 34.44$|$0.94  & \textbf{35.28}$|$\textbf{0.95} \\ \hline
			\multicolumn{2}{|l|}{Rain800} & 18.56$|$0.60 & 20.46$|$0.73 & 22.29$|$0.79 & 21.16$|$0.73 & 23.57$|$0.87 & 24.37$|$0.84 & 24.65$|$0.85 & 24.81$|$0.85 & \textbf{25.56}$|$\textbf{0.87} \\ \hline
			
	\end{tabular}}
\end{table*}

\begin{figure*}[htp!]
	\centering
	\begin{minipage}{.18\textwidth}
		\centering
		\includegraphics[width=2.7cm,height = 1.8cm]{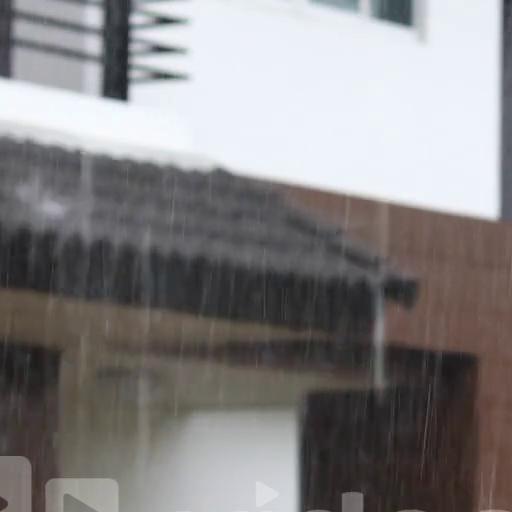}
		\vskip-5pt
		\caption*{\emph{\footnotesize{Rainy\\Image}}}
		\captionsetup{labelformat=empty}
		\captionsetup{justification=centering}
	\end{minipage}\hskip25pt
	\begin{minipage}{.18\textwidth}
		\centering
		\includegraphics[width=2.7cm,height = 1.8cm]{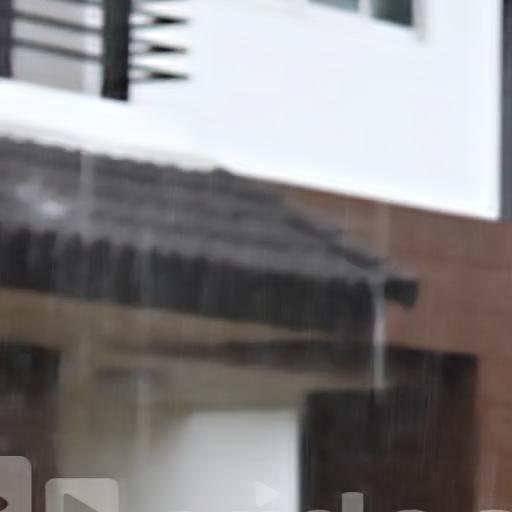}
		\vskip-5pt
		\caption*{\emph{\footnotesize{Fu~\textit{et al.}\cite{fu2017clearing}\\ (TIP'17)}}}
		\captionsetup{labelformat=empty}
		\captionsetup{justification=centering}
	\end{minipage}\hskip25pt
	\begin{minipage}{.18\textwidth}
		\centering
		\includegraphics[width=2.7cm,height = 1.8cm]{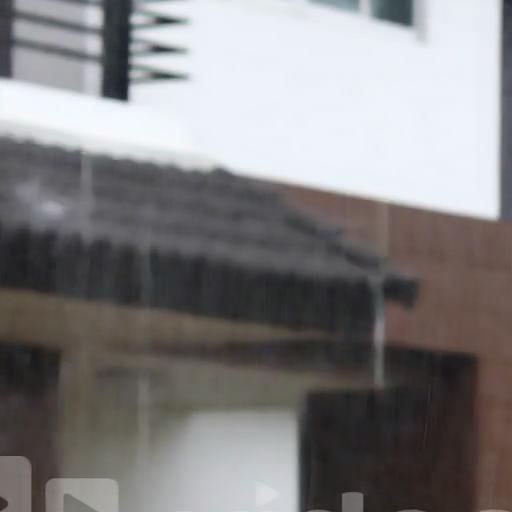}
		\vskip-5pt
		\caption*{\emph{\footnotesize{DDN}\cite{fu2017removing}\\ (CVPR'17)}}
		\captionsetup{labelformat=empty}
		\captionsetup{justification=centering}
	\end{minipage}\hskip25pt
	\begin{minipage}{.18\textwidth}
		\centering
		\includegraphics[width=2.7cm,height = 1.8cm]{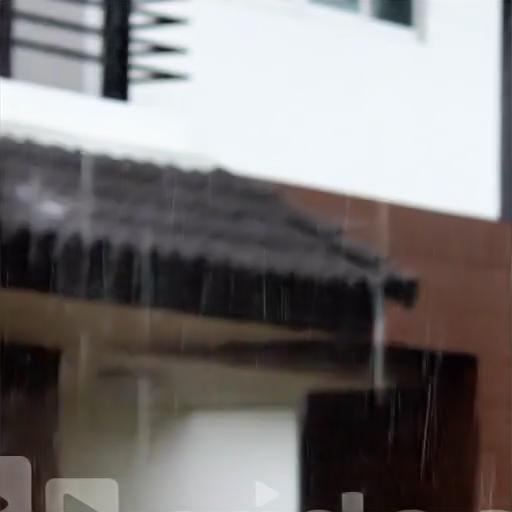}
		\vskip-5pt
		\caption*{\emph{\footnotesize{DIDMDN\cite{zhang2018density}\\ (CVPR'18)}}}
		\captionsetup{labelformat=empty}
		\captionsetup{justification=centering}
	\end{minipage}\\ \vskip5pt
	\begin{minipage}{.18\textwidth}
		\centering
		\includegraphics[width=2.7cm,height = 1.8cm]{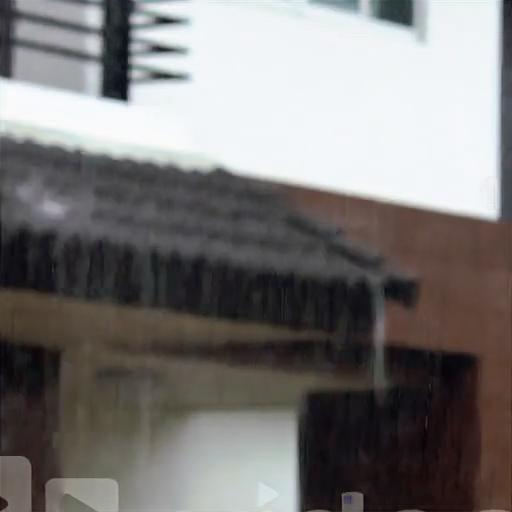}
		\vskip-5pt
		\caption*{\emph{\footnotesize{RESCAN\cite{li2018recurrent}\\ (ECCV'18)}}}
		\captionsetup{labelformat=empty}
		\captionsetup{justification=centering}
	\end{minipage}\hskip25pt
	\begin{minipage}{.18\textwidth}
		\centering
		\includegraphics[width=2.7cm,height = 1.8cm]{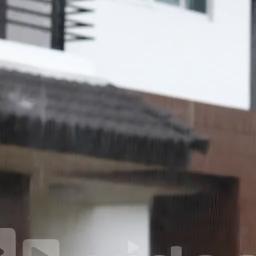}
		\vskip-5pt
		\caption*{\emph{\footnotesize{SPANet\cite{Wang_2019_CVPR}\\ (CVPR'19)}}}
		\captionsetup{labelformat=empty}
		\captionsetup{justification=centering}
	\end{minipage}\hskip25pt
	\begin{minipage}{.18\textwidth}
		\centering
		\includegraphics[width=2.7cm,height = 1.8cm]{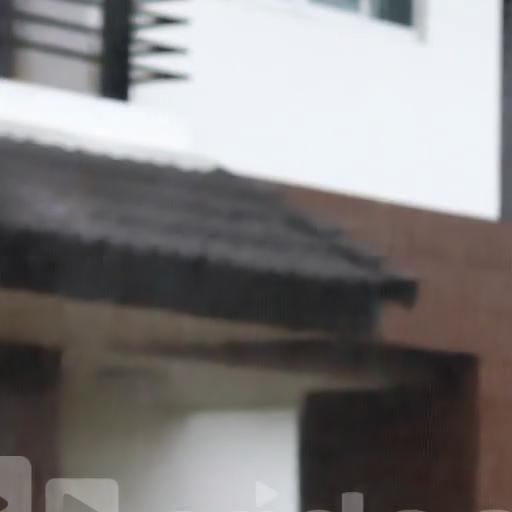}
		\vskip-5pt
		\caption*{\emph{\footnotesize{OUCD\\ (ours)}}}
		\captionsetup{labelformat=empty}
		\captionsetup{justification=centering}
	\end{minipage}\hskip25pt
	\begin{minipage}{.18\textwidth}
		\centering
		\includegraphics[width=2.7cm,height = 1.8cm]{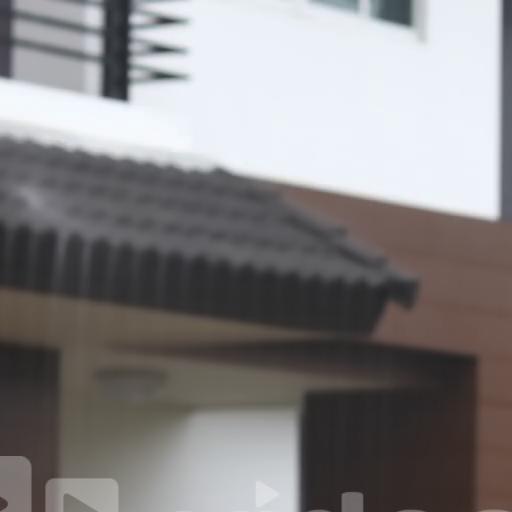}
		\vskip-5pt
		\caption*{\emph{\footnotesize{Ground-Truth\\ Image}}}
		\captionsetup{labelformat=empty}
		\captionsetup{justification=centering}
	\end{minipage}\\\vskip10pt
	\begin{minipage}{.18\textwidth}
		\centering
		\includegraphics[width=2.7cm,height = 1.8cm]{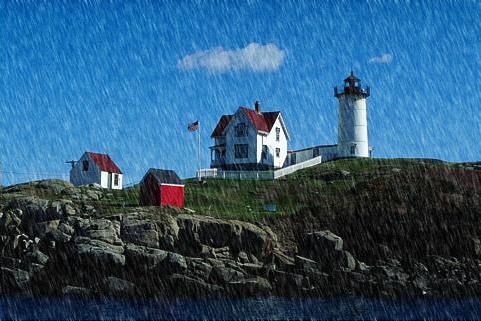}
		\vskip-5pt
		\caption*{\emph{\footnotesize{Rainy\\Image}}}
		\captionsetup{labelformat=empty}
		\captionsetup{justification=centering}
	\end{minipage}\hskip25pt
	\begin{minipage}{.18\textwidth}
		\centering
		\includegraphics[width=2.7cm,height = 1.8cm]{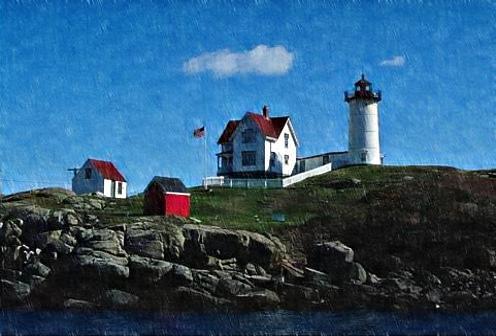}
		\vskip-5pt
		\caption*{\emph{\footnotesize{Fu~\textit{et al.}\cite{fu2017clearing}\\ (TIP'17)}}}
		\captionsetup{labelformat=empty}
		\captionsetup{justification=centering}
	\end{minipage}\hskip25pt
	\begin{minipage}{.18\textwidth}
		\centering
		\includegraphics[width=2.7cm,height = 1.8cm]{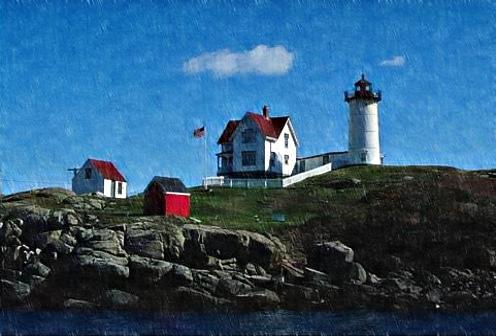}
		\vskip-5pt
		\caption*{\emph{\footnotesize{DDN}\cite{fu2017removing}\\ (CVPR'17)}}
		\captionsetup{labelformat=empty}
		\captionsetup{justification=centering}
	\end{minipage}\hskip25pt
	\begin{minipage}{.18\textwidth}
		\centering
		\includegraphics[width=2.7cm,height = 1.8cm]{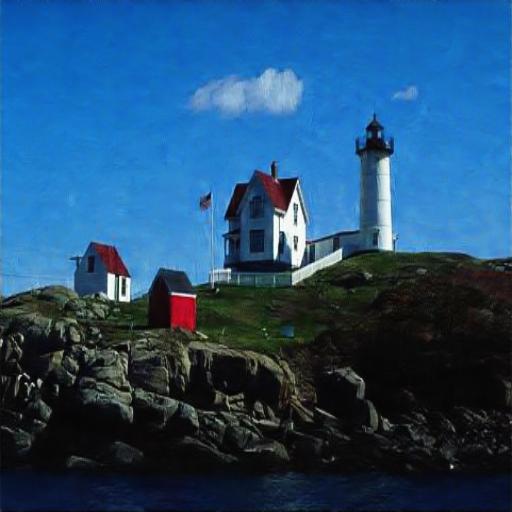}
		\vskip-5pt
		\caption*{\emph{\footnotesize{DIDMDN\cite{zhang2018density}\\ (CVPR'18)}}}
		\captionsetup{labelformat=empty}
		\captionsetup{justification=centering}
	\end{minipage}\\ \vskip5pt
	\begin{minipage}{.18\textwidth}
		\centering
		\includegraphics[width=2.7cm,height = 1.8cm]{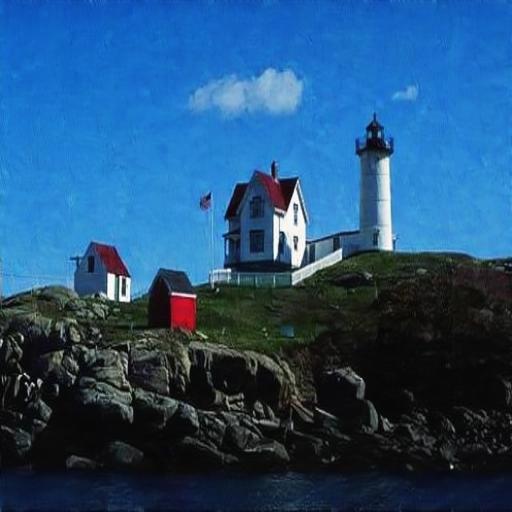}
		\vskip-5pt
		\caption*{\emph{\footnotesize{RESCAN\cite{li2018recurrent}\\ (ECCV'18)}}}
		\captionsetup{labelformat=empty}
		\captionsetup{justification=centering}
	\end{minipage}\hskip25pt
	\begin{minipage}{.18\textwidth}
		\centering
		\includegraphics[width=2.7cm,height = 1.8cm]{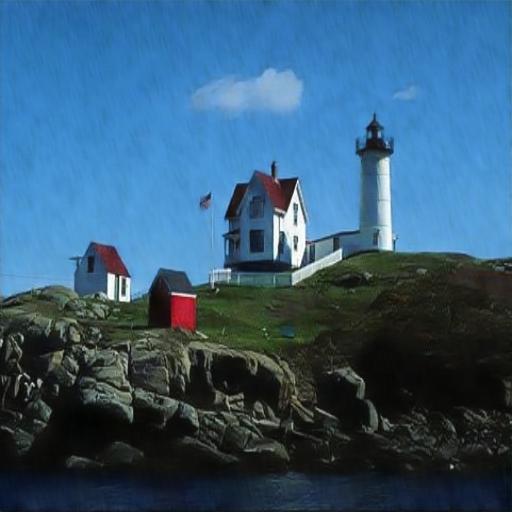}
		\vskip-5pt
		\caption*{\emph{\footnotesize{SPANet\cite{Wang_2019_CVPR}\\ (CVPR'19)}}}
		\captionsetup{labelformat=empty}
		\captionsetup{justification=centering}
	\end{minipage}\hskip25pt
	\begin{minipage}{.18\textwidth}
		\centering
		\includegraphics[width=2.7cm,height = 1.8cm]{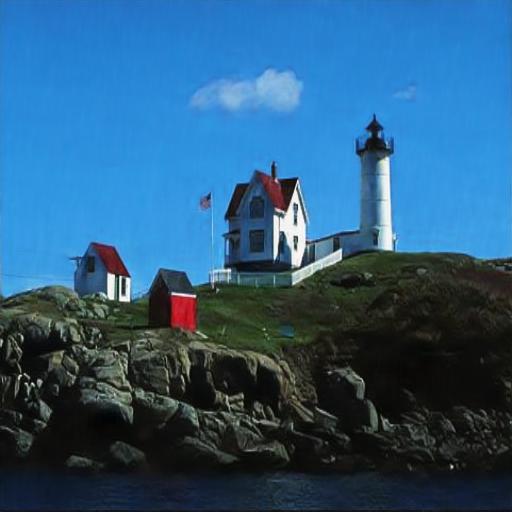}
		\vskip-5pt
		\caption*{\emph{\footnotesize{OUCD\\ (ours)}}}
		\captionsetup{labelformat=empty}
		\captionsetup{justification=centering}
	\end{minipage}\hskip25pt
	\begin{minipage}{.18\textwidth}
		\centering
		\includegraphics[width=2.7cm,height = 1.8cm]{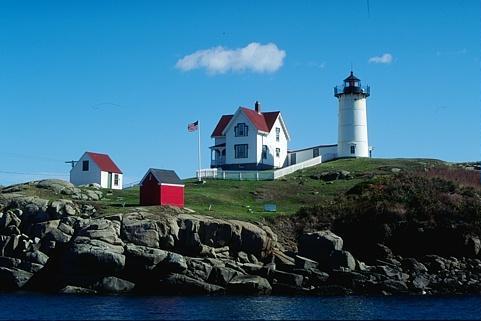}
		\vskip-5pt
		\caption*{\emph{\footnotesize{Ground-Truth\\ Image}}}
		\captionsetup{labelformat=empty}
		\captionsetup{justification=centering}
	\end{minipage}\\ \vskip5pt
	\begin{minipage}{.18\textwidth}
		\centering
		\includegraphics[width=2.7cm,height = 1.8cm]{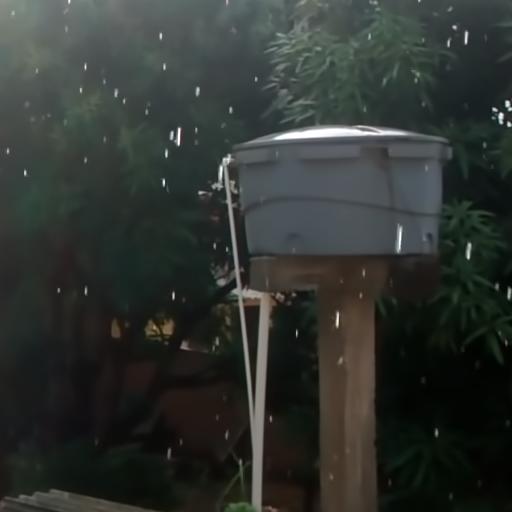}
		\vskip-5pt
		\caption*{\emph{\footnotesize{Rainy\\Image}}}
		\captionsetup{labelformat=empty}
		\captionsetup{justification=centering}
	\end{minipage}\hskip25pt
	\begin{minipage}{.18\textwidth}
		\centering
		\includegraphics[width=2.7cm,height = 1.8cm]{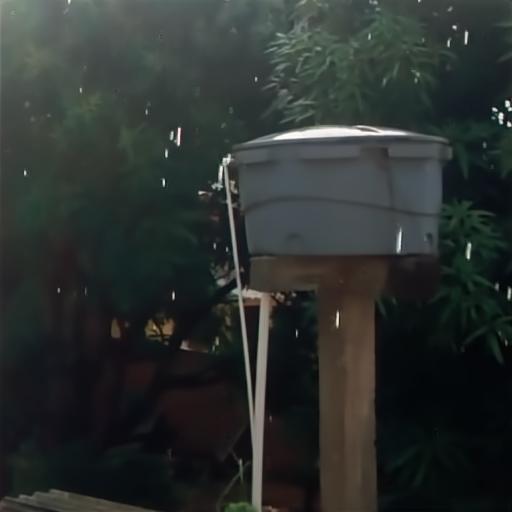}
		\vskip-5pt
		\caption*{\emph{\footnotesize{Fu~\textit{et al.}\cite{fu2017clearing}\\ (TIP'17)}}}
		\captionsetup{labelformat=empty}
		\captionsetup{justification=centering}
	\end{minipage}\hskip25pt
	\begin{minipage}{.18\textwidth}
		\centering
		\includegraphics[width=2.7cm,height = 1.8cm]{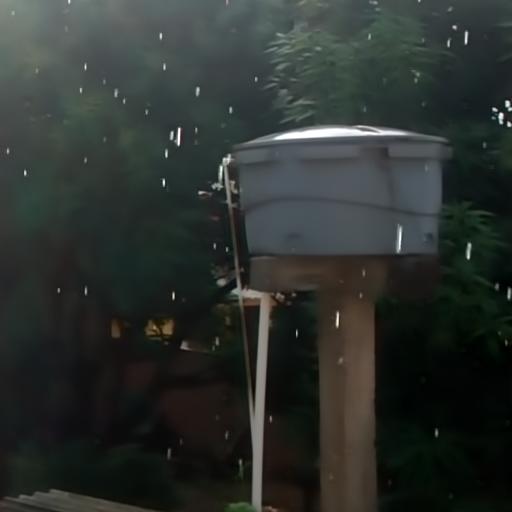}
		\vskip-5pt
		\caption*{\emph{\footnotesize{DDN}\cite{fu2017removing}\\ (CVPR'17)}}
		\captionsetup{labelformat=empty}
		\captionsetup{justification=centering}
	\end{minipage}\hskip25pt
	\begin{minipage}{.18\textwidth}
		\centering
		\includegraphics[width=2.7cm,height = 1.8cm]{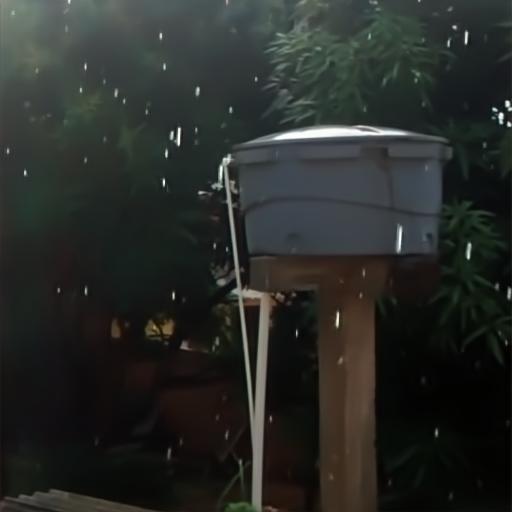}
		\vskip-5pt
		\caption*{\emph{\footnotesize{DIDMDN\cite{zhang2018density}\\ (CVPR'18)}}}
		\captionsetup{labelformat=empty}
		\captionsetup{justification=centering}
	\end{minipage}\\ \vskip5pt
	\begin{minipage}{.18\textwidth}
		\centering
		\includegraphics[width=2.7cm,height = 1.8cm]{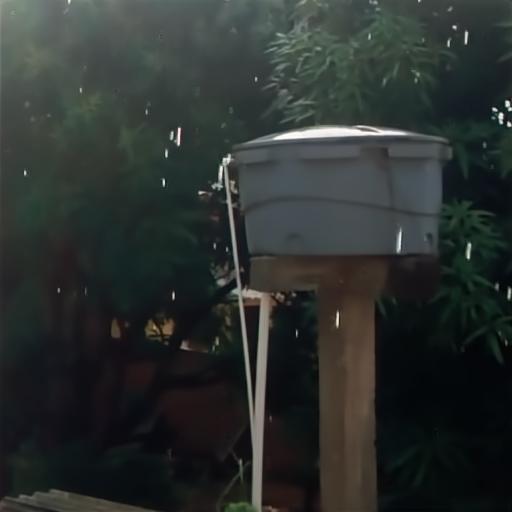}
		\vskip-5pt
		\caption*{\emph{\footnotesize{RESCAN\cite{li2018recurrent}\\ (ECCV'18)}}}
		\captionsetup{labelformat=empty}
		\captionsetup{justification=centering}
	\end{minipage}\hskip25pt
	\begin{minipage}{.18\textwidth}
		\centering
		\includegraphics[width=2.7cm,height = 1.8cm]{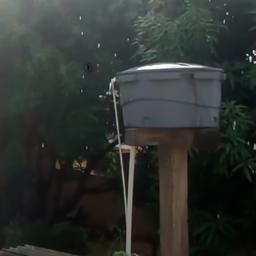}
		\vskip-5pt
		\caption*{\emph{\footnotesize{SPANet\cite{Wang_2019_CVPR}\\ (CVPR'19)}}}
		\captionsetup{labelformat=empty}
		\captionsetup{justification=centering}
	\end{minipage}\hskip25pt
	\begin{minipage}{.18\textwidth}
		\centering
		\includegraphics[width=2.7cm,height = 1.8cm]{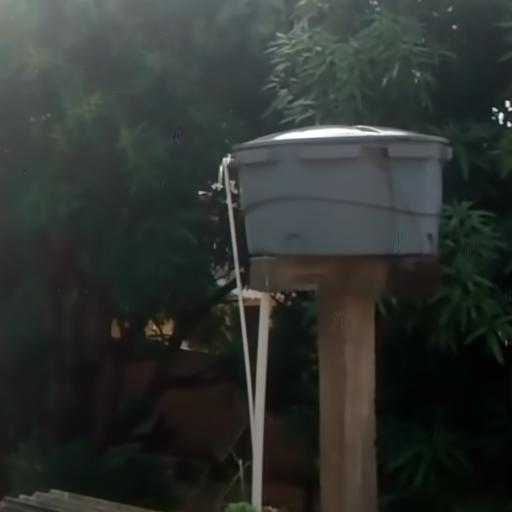}
		\vskip-5pt
		\caption*{\emph{\footnotesize{OUCD\\ (ours)}}}
		\captionsetup{labelformat=empty}
		\captionsetup{justification=centering}
	\end{minipage}\hskip25pt
	\begin{minipage}{.18\textwidth}
		\centering
		\includegraphics[width=2.7cm,height = 1.8cm]{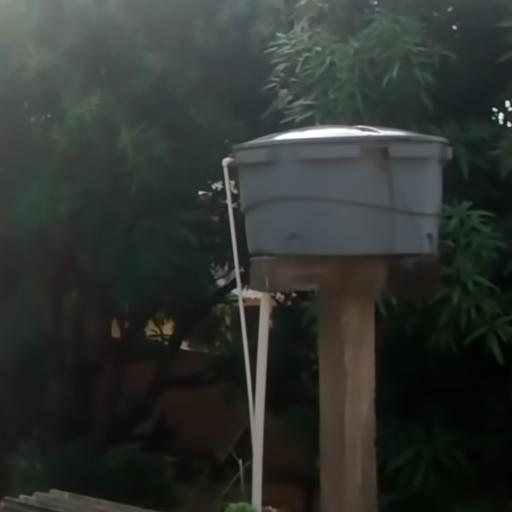}
		\vskip-5pt
		\caption*{\emph{\footnotesize{Ground-Truth\\ Image}}}
		\captionsetup{labelformat=empty}
		\captionsetup{justification=centering}
	\end{minipage}\\\vskip10pt
	
	\vskip2pt
	\caption{Qualitative comparisons of OUCD with recent methods on deraining of the test datasets of SPANet and Rain800. In the first image, most of the other methods fail to remove small rain streaks near the roof of the image whereas our proposed method produces a very clear image without any rain streaks. In the second image, though the other methods are able to remove the rain streaks, they fail to produce a proper reconstruction of the background. It can be seen that our method's output has a clearer background and is closer to the ground truth than the others. In the third image, all the previous methods fail to remove the tiny rain drops whereas OUCD removes all the rain drops.}
	\label{Fig:exp3}
\end{figure*}

\begin{figure*}[htbp!]
	\centering\hskip-40pt
	\begin{minipage}{.18\textwidth}
		\centering
		\includegraphics[width=3.6cm,height = 2.4cm]{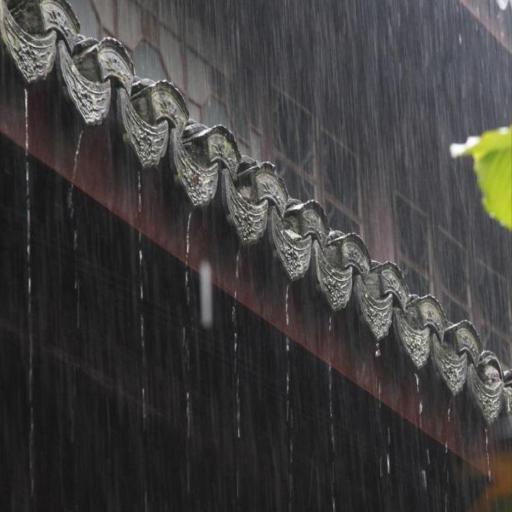}
		\vskip-5pt
		\caption*{\emph{\footnotesize{Rainy\\Image}}}
		\captionsetup{labelformat=empty}
		\captionsetup{justification=centering}
	\end{minipage}\hskip45pt
	\begin{minipage}{.18\textwidth}
		\centering
		\includegraphics[width=3.6cm,height = 2.4cm]{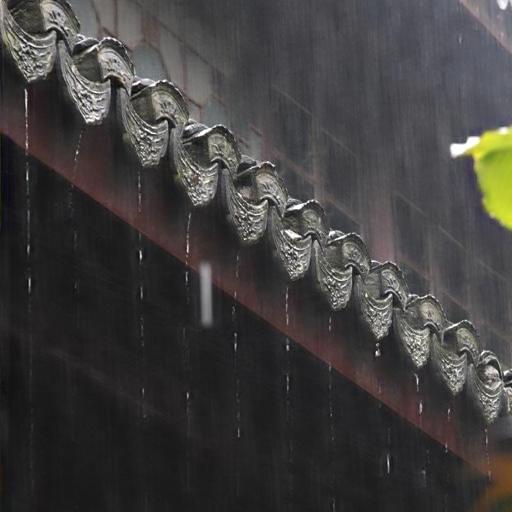}
		\vskip-5pt
		\caption*{\emph{\footnotesize{DDN}\cite{fu2017removing}\\ (CVPR'17)}}
		\captionsetup{labelformat=empty}
		\captionsetup{justification=centering}
	\end{minipage}\hskip45pt
	\begin{minipage}{.18\textwidth}
		\centering
		\includegraphics[width=3.6cm,height = 2.4cm]{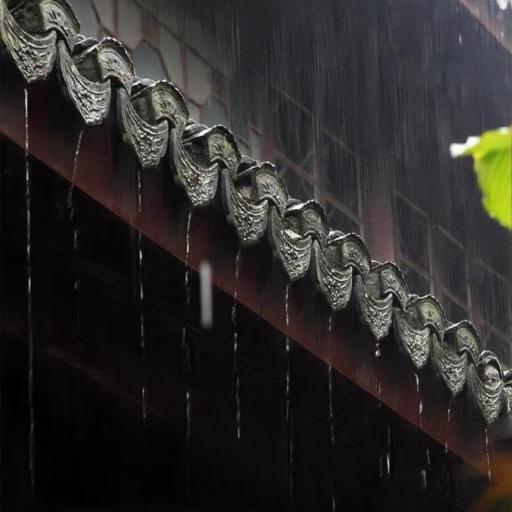}
		\vskip-5pt
		\caption*{\emph{\footnotesize{DIDMDN\cite{zhang2018density}\\ (CVPR'18)}}}
		\captionsetup{labelformat=empty}
		\captionsetup{justification=centering}
	\end{minipage}\\ \vskip5pt \hskip-40pt
	\begin{minipage}{.18\textwidth}
		\centering
		\includegraphics[width=3.6cm,height = 2.4cm]{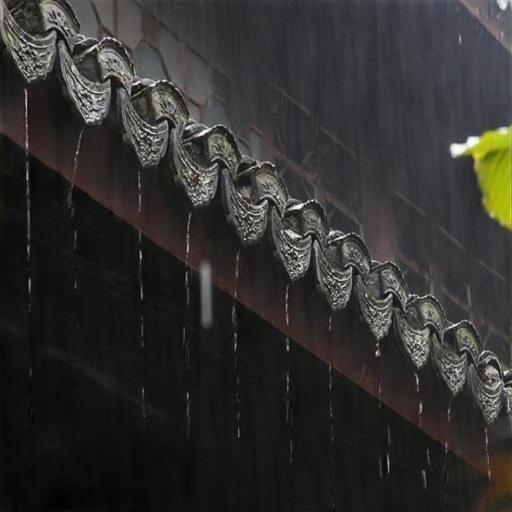}
		\vskip-5pt
		\caption*{\emph{\footnotesize{RESCAN\cite{li2018recurrent}\\ (ECCV'18)}}}
		\captionsetup{labelformat=empty}
		\captionsetup{justification=centering}
	\end{minipage}\hskip45pt
	\begin{minipage}{.18\textwidth}
		\centering
		\includegraphics[width=3.6cm,height = 2.4cm]{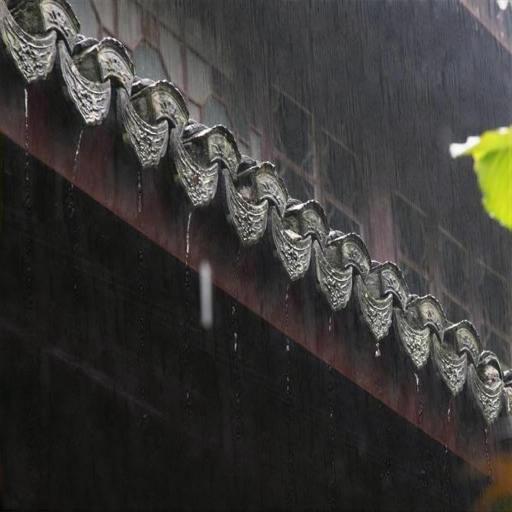}
		\vskip-5pt
		\caption*{\emph{\footnotesize{SPANet\cite{Wang_2019_CVPR}\\ (CVPR'19)}}}
		\captionsetup{labelformat=empty}
		\captionsetup{justification=centering}
	\end{minipage}\hskip45pt
	\begin{minipage}{.18\textwidth}
		\centering
		\includegraphics[width=3.6cm,height = 2.4cm]{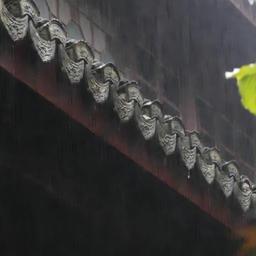}
		\vskip-5pt
		\caption*{\emph{\footnotesize{OUCD\\ (ours)}}}
		\captionsetup{labelformat=empty}
		\captionsetup{justification=centering}
	\end{minipage}\\ \vskip10pt \hskip-40pt 
	\begin{minipage}{.18\textwidth}
		\centering
		\includegraphics[width=3.6cm,height = 2.4cm]{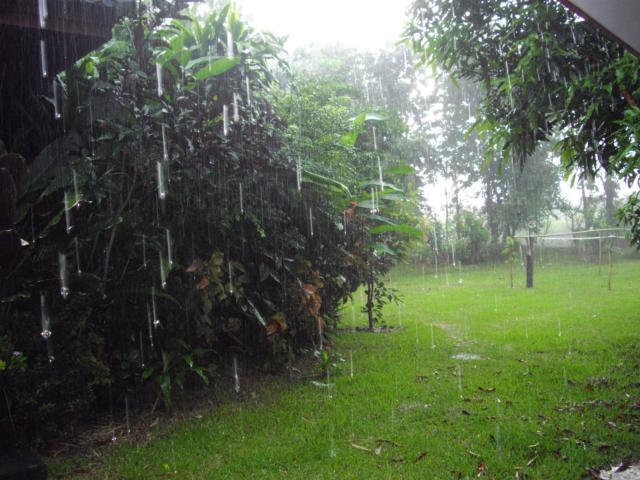}
		\vskip-5pt
		\caption*{\emph{\footnotesize{Rainy\\Image}}}
		\captionsetup{labelformat=empty}
		\captionsetup{justification=centering}
	\end{minipage}\hskip45pt
	\begin{minipage}{.18\textwidth}
		\centering
		\includegraphics[width=3.6cm,height = 2.4cm]{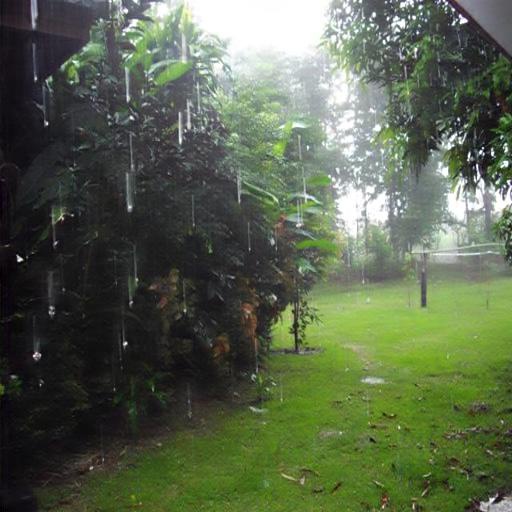}
		\vskip-5pt
		\caption*{\emph{\footnotesize{DDN}\cite{fu2017removing}\\ (CVPR'17)}}
		\captionsetup{labelformat=empty}
		\captionsetup{justification=centering}
	\end{minipage}\hskip45pt
	\begin{minipage}{.18\textwidth}
		\centering
		\includegraphics[width=3.6cm,height = 2.4cm]{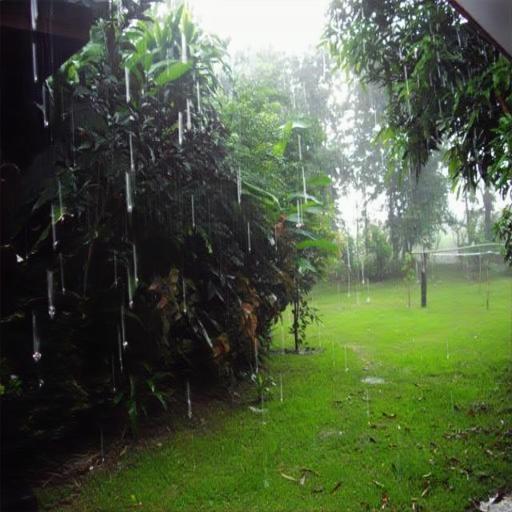}
		\vskip-5pt
		\caption*{\emph{\footnotesize{DIDMDN\cite{zhang2018density}\\ (CVPR'18)}}}
		\captionsetup{labelformat=empty}
		\captionsetup{justification=centering}
	\end{minipage}\\ \hskip-40pt
	\begin{minipage}{.18\textwidth}
		\centering
		\includegraphics[width=3.6cm,height = 2.4cm]{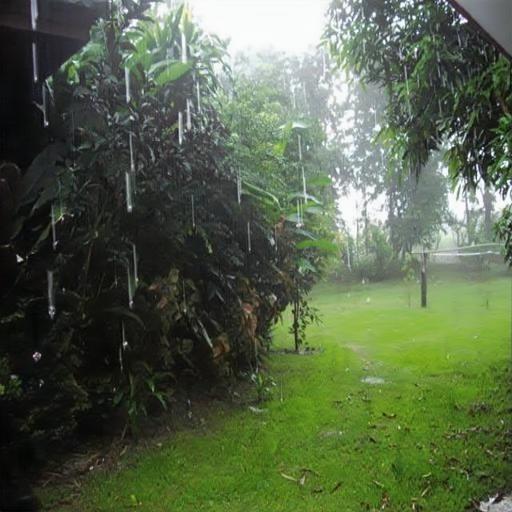}
		\vskip-5pt
		\caption*{\emph{\footnotesize{RESCAN\cite{li2018recurrent}\\ (ECCV'18)}}}
		\captionsetup{labelformat=empty}
		\captionsetup{justification=centering}
	\end{minipage}\hskip45pt
	\begin{minipage}{.18\textwidth}
		\centering
		\includegraphics[width=3.6cm,height = 2.4cm]{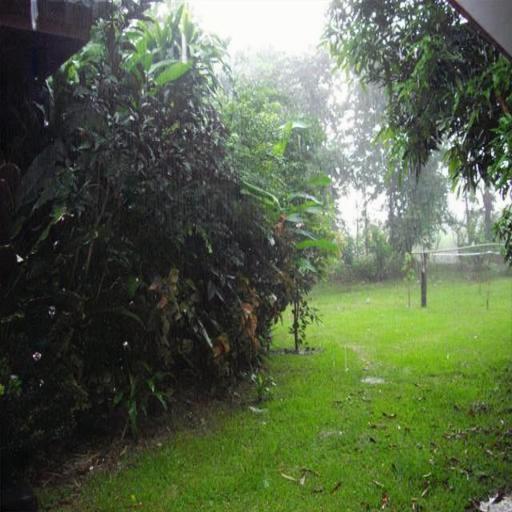}
		\vskip-5pt
		\caption*{\emph{\footnotesize{SPANet\cite{Wang_2019_CVPR}\\ (CVPR'19)}}}
		\captionsetup{labelformat=empty}
		\captionsetup{justification=centering}
	\end{minipage}\hskip45pt
	\begin{minipage}{.18\textwidth}
		\centering
		\includegraphics[width=3.6cm,height = 2.4cm]{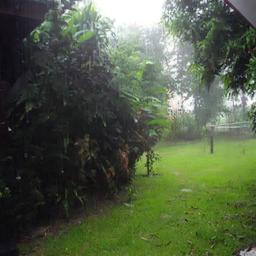}
		\vskip-5pt
		\caption*{\emph{\footnotesize{OUCD \\ (ours)}}}
		\captionsetup{labelformat=empty}
		\captionsetup{justification=centering}
	\end{minipage}
	\vskip-5pt
	\caption{Qualitative comparisons of OUCD with recent methods on deraining for real-world rainy images. In both of the illustrated examples, it can be seen that OUCD gives better derained images than the other methods. In the first image, even though the other methods remove the dense streaks present in the top-right part of the image, they fail to remove the rain streaks of different patterns found near the roof. Our method removes rain streak of any pattern effectively. Also in the second image all the other methods remove the rain streaks which are far off effectively but fail to remove the rain streaks which are closer to the camera. As OUCD has two encoders, where one works on local features and another on global features, it can be seen that OUCD removes the rain streaks which are far away (small drops) and which are closer to camera (large drops) from the camera efficiently. Note that these are real images and do not come with ground truths. }
	\label{Fig:exp4}
\end{figure*}	

\subsection{Quantitative Results}
We compare OUCD network quantitatively against the state-of-the-art methods using the synthetic and real test set images mentioned earlier. Table~\ref{experiments} shows the quantitative comparisons. We can clearly observe that the OUCD network outperforms the current state-of-the-art methods. On heavy rain synthetic datasets like DIDMDN (both Test-1 and Test-2) and Rain800, our OUCD Network performs approximately $\sim 1$dB better than the state-of-the-art methods in terms of PSNR. For low rain synthetic datasets like Rain12, OUCD also performs better than the current methods by $\sim 1.1$dB. It is interesting to observe that OUCD outperforms the state-of-the-art methods by $\sim1.25$dB on real rainy dataset SPANet \cite{Wang_2019_CVPR}.

Note that our work proposes an architecture change that focuses on deraining.  Our network can be used in place of U-Net or residual architectures along with a confidence-based loss functions \cite{yasarla2019uncertainty} or progressive deraining \cite{ren2019progressive} to even further improve the performance.  In this paper, we have shown that we get significant performance improvements over the recent methods with just the architecture change, while using only generic loss functions such as L2-loss and the perceptual loss.

\subsection{Qualitative Results}	
We compare OUCD qualitatively against  state-of-the-art methods using the SPANet and Rain800 test sets, and real rainy images shared by the authors of \cite{Wang_2019_CVPR,zhang2018density}. It can be clearly observed from the Fig.~\ref{Fig:exp3} that OUCD performs visually better than the state-of-the-art methods on the SPANet and Rain800 test sets. From  Fig.~\ref{Fig:exp3}, we clearly see that the state-of-the-art methods which are based on Res-Net or U-Net architectures are not able remove the rain streaks on the tiles of the roof in the first image, and rain streaks on the tower and the house roof top in the second image. For example, DIDMDN \cite{zhang2018density}, which is based on U-Net architecture, is not able to remove streaks from tiles on the roof in the first image, and rain streaks on clouds, roof tops and tower in the second image. This is due to the fact that U-Net based architectures do not focus much on local features as much as it should. DDN \cite{fu2017removing}, RESCAN \cite{li2018recurrent}, SPANet \cite{Wang_2019_CVPR}, which are based on Res-Net, show a similar performance as they are not able to remove rain streaks where the removal of rain streaks is hard around the textured regions like tiles on the roof, near clouds and top of the tower. On the other hand, our OUCD network which focuses on features both locally and globally while deraining the image, produces sharp and visually pleasing derained images.

\begin{figure*}[htp!]
	\centering\hskip-40pt
	\begin{minipage}{.18\textwidth}
		\centering
		\includegraphics[width=3.6cm,height = 2.4cm]{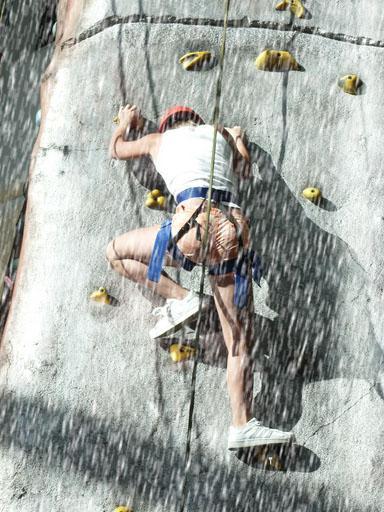}
		\vskip-5pt
		\caption*{\emph{\scriptsize{Rainy Image}}}
		\captionsetup{labelformat=empty}
		\captionsetup{justification=centering}
	\end{minipage}\hskip45pt
	\begin{minipage}{.18\textwidth}
		\centering
		\includegraphics[width=3.6cm,height = 2.4cm]{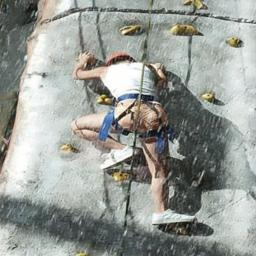}
		\vskip-5pt
		\caption*{\emph{\scriptsize{Under-complete}}}
		\captionsetup{labelformat=empty}
		\captionsetup{justification=centering}
	\end{minipage}\hskip45pt
	\begin{minipage}{.18\textwidth}
		\centering
		\includegraphics[width=3.6cm,height = 2.4cm]{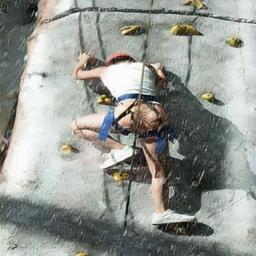}
		\vskip-5pt
		\caption*{\emph{\scriptsize{Over-complete}}}
		\captionsetup{labelformat=empty}
		\captionsetup{justification=centering}
	\end{minipage}\\ \vskip10pt \hskip-40pt
	\begin{minipage}{.18\textwidth}
		\centering
		\includegraphics[width=3.6cm,height = 2.4cm]{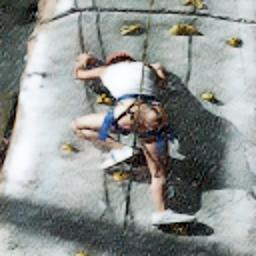}
		\vskip-5pt
		\caption*{\emph{\scriptsize{Combined at last}}}
		\captionsetup{labelformat=empty}
		\captionsetup{justification=centering}
	\end{minipage}\hskip45pt
	\begin{minipage}{.18\textwidth}
		\centering
		\includegraphics[width=3.6cm,height = 2.4cm]{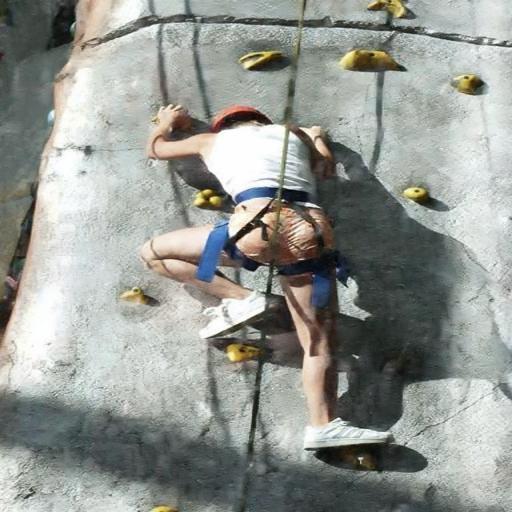}
		\vskip-5pt
		\caption*{\emph{\scriptsize{OUCD (ours)}}}
		\captionsetup{labelformat=empty}
		\captionsetup{justification=centering}
	\end{minipage}\hskip45pt
	\begin{minipage}{.18\textwidth}
		\centering
		\includegraphics[width=3.6cm,height = 2.4cm]{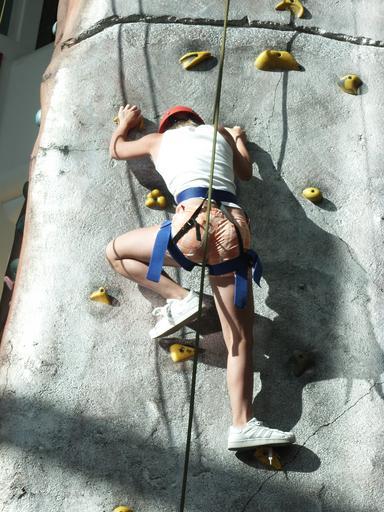}
		\vskip-5pt
		\caption*{\emph{\scriptsize{Ground-Truth}}}
		\captionsetup{labelformat=empty}
		\captionsetup{justification=centering}
	\end{minipage}\\ \vskip10pt \hskip-40pt 
	\vskip-20pt
	\caption{Qualitative comparisons corresponding to ablation study. The undercomplete network essentially removes relatively larger rain streaks and the overcomplete network removes smaller streaks. When the features from both the networks are combined in a conventional way (at last), they produce a better derained image when compared to separate networks. Our proposed method OUCD which combines the features using MSFF produces a better derained image compared to conventional fusion. }
	\label{Fig:exp5}
\end{figure*}

\begin{table}[htp!]
	\centering
	\caption{PSNR and SSIM (PSNR$|$SSIM) quantitative comparisons corresponding to ablation study.}
	\label{ablation_quan}
	\resizebox{0.48\textwidth}{!}{
		\begin{tabular}{|l|c|c|}
			\hline
			Method & Rain800 & Rain12 \\ \hline
			under complete UNet & 22.99$|$0.74 & 25.50$|$0.77 \\ \hline
			Overcomplete UNet & 23.74$|$0.80 & 29.26$|$ 0.89 \\ \hline
			OUCD w/o MSFF block & 24.94$|$0.82 & 31.01$|$0.91 \\ \hline
			OUCD w/ MSFF block & \textbf{25.56}$|$\textbf{0.87} & \textbf{35.28}$|$\textbf{0.95} \\ \hline
	\end{tabular}}
\end{table}

We observe similar trend in visual quality of these state-of-the-art methods on real rainy images provided by the authors of  \cite{Wang_2019_CVPR,zhang2018density}. From Fig.~\ref{Fig:exp4} we can clearly observe that deraining methods \cite{fu2017removing,zhang2018density,li2018recurrent} are not able remove rain streaks on the roof in the first image, and rain streaks on the tree in the second image. Even though SPANet~\cite{Wang_2019_CVPR} addresses the deraining problem using spatially attentive modules, the predicted derained image has small rain streaks present on the roof tile in the first image, and leaves of the tree in the second image of Fig.~\ref{Fig:exp4}. Our method  produces sharper derained images even on the real rainy images as shown in the Fig.~\ref{Fig:exp4}. For example, tiles of the roof are more clear and sharp in the first derained image, and there are no rain streaks on the leaves of the tree in the second derained image of OUCD in Fig.~\ref{Fig:exp4}.

\subsection{Ablation Study}
We perform an ablation study to show the impact of each block in our proposed method. We start our experiment with just a generic undercomplete architecture which is basically a U-Net. Then we test with an overcomplete architecture which focuses heavily on local features. From Table \ref{ablation_quan}, it can be seen that the overcomplete architecture gives a better performance than the undercomplete network. It can be noted here that the number of blocks and parameters used in the overcomplete architecture is relatively lesser.  We then check the performance when both the overcomplete and undercomplete networks are trained in parallel and fused at the last layer (OUCD w/o MSFF block). This gives a boost in performance as the fused network has two branches where one solely focuses on low-level features while the other branch captures both low and high-level features thus the overall network does not lose out on global features altogether unlike the overcomplete architecture. Next, we experiment OUCD network while using our proposed MSFF block. Using the MSFF block to transfer the feature maps from the overcomplete branch to the undercomplete branch helps the undercomplete branch  to train better as they now have better quality feature maps in the early layers which can facilitate better feature capturing in the deeper layers. This can be seen from Table \ref{ablation_quan} as we get a significant improvements when compared to just fusing the networks at the last layer. Figure \ref{Fig:exp5} illustrates qualitative results corresponding to the ablation study.

\section{Conclusion}
We propose using overcomplete convolutional deep networks for the task of image deraining. We showed that using overcomplete architectures help to restrict the receptive field size of the filters in the deeper layers. This helps the filters in deep layers to capture more low-level information and finer details than the generic ``encoder-decoder" architectures or residual networks. Focussing on low-level features makes much more sense in the task of deraining as the rain streaks are generally of a small size in the rainy image. We proposed an Over-and-Under Complete Derain (OUCD) Network which takes advantage of both local and global features extracted from its overcomplete and undercomplete branches, respectively. We use a special Multi Scale Feature Fusion (MSFF) block to efficiently transfer feature maps from the overcomplete branch to the undercomplete branch so as to aid the undercomplete network's training  with better low-level feature maps extracted in the overcomplete branch.   We show that the filters in the overcomplete branch of our proposed network captures even the tiniest streaks of the rain precisely aiding in deraining, and the undercomplete branch captures objects and other structures in the image  aiding in proper reconstruction.   We tested our method on both real and synthetic rain datasets availble in the literature where we achieve better performance than many recent state-of-the-art methods. We achieve a significant performance boost with just an architecture change while using generic loss functions like L2 loss and the perceptual loss. We also perform an ablation study to show the impact of each  block that we propose in this work.  
\appendices
\section{Network Architecture}

Tables \ref{tab1} and \ref{tab2} show the configuration of the undercomplete and overcomplete branches of our OUCD Network. Table \ref{Table:run_time} shows the average inference time comparison between our proposed methods other state of the art methods. Total number of parameters in OCUD network are $1.1 \times 10^7$.
 
\section{Receptive field size}
 \begin{figure}[htbp]
 	\centering
 	\includegraphics[width=.8\linewidth]{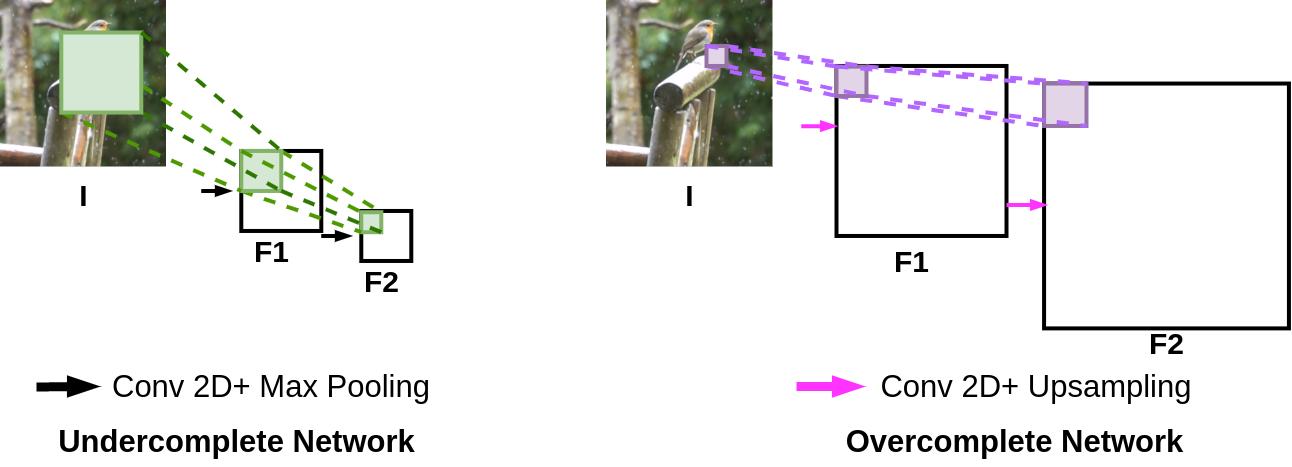}\\
 	\caption{Explanation for receptive field change difference between undercomplete network and overcomplete network.}
 	\label{explana}
 \end{figure}
 Let $I$ be the input image, $F_1$ and $F_2$ be the feature maps extracted from the conv blocks 1 and 2, respectively. The max-pooling layer present in these conv blocks of the U-Net is the main reason why receptive field is large in the successive layers. Let the initial receptive field of the conv filter be $k \times k$ on the image. The receptive field size change due to max-pooling layer is dependent on two variables- pooling coefficient and stride of the pooling filter. For convenience, the pooling coefficient and stride is both set as 2 in UNet. Considering this configuration, the receptive field of conv block 2 (to which $F_1$ is forwarded) on the input image would be $ 2 \times k \times 2 \times k$. Similarly, the receptive field of conv block 3 (to which $F_2$ is forwarded) would be $ 4 \times k \times 4 \times k$. This increase in receptive field can be generalized for $i^{th}$ layer in the UNet as follows:
 
 \[ RF (w.r.t \; I) =  2^{2*(i-1)} \times k \times k \]
 
 Now for our proposed overcomplete network, we have upsampling layer of coefficient 2 in our conv blocks replacing the max-pooling layer. As the upsampling layer actually works exactly opposite to that of max-pooling layer, the receptive field of conv bock 2 on the input image now would be $ \frac{1}{2} \times k \times \frac{1}{2} \times k$. Similarly, the receptive field of conv block 3  now would be $ \frac{1}{4} \times k \times \frac{1}{4} \times k$. This increase in receptive field can be generalized for $i^{th}$ layer in the overcomplete branch as follows:  
 
 \[ RF (w.r.t \; I) =  (\frac{1}{2})^{2*(i-1)} \times k \times k \]
 
 Note that the above calculations are based on a couple of assumptions. We assume that the pooling coefficient and stride is both set as 2 in both overcomplete and undercomplete network. Also we consider that the receptive field change caused by the conv layer in both undercomplete and overcomplete networks would be the same and do not consider in our calculations. This can be justified as we have maintained the conv kernel size to $3 \times 3$ with stride 1 and padding 1 throughout our network and this setting does not actually affect the receptive as much as max-pooling or upsampling layer does. The above explanations are illustrated in Fig \ref{explana}.
\begin{table}[htp!]
	\begin{center}
		\centering
		\caption{Configuration of blocks in the overcomplete branch of OUCD.}
		
		\resizebox{0.48\textwidth}{!}{
			\begin{tabular}{c|c|c|c|c|c|c}
				\hline \hline
				Block name                          & Layer     & Kernel size/Scale Factor& Filters & Padding & Input size & Output size \\ \hline
				\multirow{9}{*}{Encoder}           & Conv1     & 3 $\times$ 3       & 32      & 1          & 3 $\times$ H $\times$ W  & 32 $\times$ H $\times$ W  \\ \cline{2-7} 
				& Upsampling     & 2 $\times$ 2       & -      & -          & 32 $\times$ H $\times$ W & 32 $\times$ 2H $\times$ 2W  \\ \cline{2-7} 
				& ReLU     & -      & -       & -          & 32 $\times$ 2H $\times$ 2W & 32 $\times$ 2H $\times$ 2W   \\ \cline{2-7}
				& Conv2    & 3 $\times$ 3       & 64      & 1          & 32 $\times$ 2H $\times$ 2W  & 64 $\times$ 2H $\times$ 2W  \\ \cline{2-7} 
				& Upsampling     & 2 $\times$ 2       & -      & -          & 64 $\times$ 2H $\times$ 2W & 64 $\times$ 4H $\times$ 4W  \\ \cline{2-7} 
				& ReLU     & -       & -       & -          & 64 $\times$ 4H $\times$ 4W & 64 $\times$ 4H $\times$ 4W   \\ \cline{2-7}
				& Conv3     & 3 $\times$ 3        & 128      & 1          & 64 $\times$ 4H $\times$ 4W  & 128 $\times$ 4H $\times$ 4W  \\ \cline{2-7} 
				& Upsampling     & 2 $\times$ 2       & -      & -          & 128 $\times$ 4H $\times$ 4W & 2C $\times$ 8H $\times$ 8W  \\ \cline{2-7} 
				& ReLU     & -       & -       & -          & 128 $\times$ 8H $\times$ 8W & 128 $\times$ 8H $\times$ 8W   \\ \hline					
				\multirow{9}{*}{Decoder}           & Conv1     & 3 $\times$ 3       & 128      & 1          & 128 $\times$ 8H $\times$ 8W  & 128 $\times$ 8H $\times$ 8W  \\ \cline{2-7} 
				& Max-Pooling     & 2 $\times$ 2       & -      & -          & 128 $\times$ 8H $\times$ 8W & 128 $\times$ 4H $\times$ 4W  \\ \cline{2-7} 
				& ReLU     & -      & -       & -          & 128 $\times$ 4H $\times$ 4W & 128 $\times$ 4H $\times$ 4W   \\ \cline{2-7}
				& Conv2    & 3 $\times$ 3       & 64      & 1          & 128 $\times$ 4H $\times$ 4W  & 128 $\times$ 4H $\times$ 4W  \\ \cline{2-7} 
				& Max-Pooling      & 2 $\times$ 2       & -      & -          & 64 $\times$ 4H $\times$ 4W & 64 $\times$ 2H $\times$ 2W  \\ \cline{2-7} 
				& ReLU     & -       & -       & -          & 64 $\times$ 2H $\times$ 2W & 64 $\times$ 2H $\times$ 2W   \\ \cline{2-7}
				& Conv3     & 3 $\times$ 3        & 32      & 1          & 64 $\times$ 2H $\times$ 2W  & 32 $\times$ 2H $\times$ 2W  \\ \cline{2-7} 
				& Max-Pooling      & 2 $\times$ 2       & -      & -          & 32 $\times$ 2H $\times$ 2W & 32 $\times$ H $\times$ W  \\ \cline{2-7} 
				& ReLU     & -       & -       & -          & 32 $\times$ H $\times$ W & 32 $\times$ H $\times$ W   \\ \hline

			\end{tabular}
		}
		\label{tab1}
	\end{center}
\end{table}

\begin{table}[htp!]
	\begin{center}
		\centering
		\caption{Configuration of the undercomplete branch of OUCD.}
		\label{tab2}
		\resizebox{0.5\textwidth}{!}{
			\begin{tabular}{c|c|c|c|c|c|c}
				\hline \hline
				Block name                          & Layer     & Kernel size/Scale Factor& Filters & Padding & Input size & Output size \\ \hline

				\multirow{15}{*}{Encoder}           & Conv1     & 3 $\times$ 3       & 32      & 1          & 3 $\times$ H $\times$ W  & 32 $\times$ H $\times$ W  \\ \cline{2-7} 
				& MaxPooling     & 2 $\times$ 2       & -      & -          & 32 $\times$ H $\times$ W & 32 $\times$ H/2 $\times$ W/2  \\ \cline{2-7} 
				& ReLU     & -      & -       & -          & 32 $\times$ H/2 $\times$ W/2 & 32 $\times$ H/2 $\times$ W/2   \\ \cline{2-7}
				& Conv2    & 3 $\times$ 3       & 64      & 1          & 32 $\times$ H/2 $\times$ W/2  & 64 $\times$ H/2 $\times$ W/2  \\ \cline{2-7} 
				& MaxPooling     & 2 $\times$ 2       & -      & -          & 64 $\times$ H/2 $\times$ W/2 & 64 $\times$ H/4 $\times$ W/4  \\ \cline{2-7} 
				& ReLU     & -       & -       & -          & 64 $\times$ H/4 $\times$ W/4 & 64 $\times$ H/4 $\times$ W/4   \\ \cline{2-7}
				& Conv3     & 3 $\times$ 3        & 128      & 1          & 64 $\times$ H/4 $\times$ W/4  & 128 $\times$ H/4 $\times$ W/4  \\ \cline{2-7} 
				& MaxPooling     & 2 $\times$ 2       & -      & -          & 128 $\times$ H/4 $\times$ W/4 & 128 $\times$ H/8 $\times$ W/8  \\ \cline{2-7} 
				& ReLU     & -       & -       & -          & 128 $\times$ H/8 $\times$ W/8 & 128 $\times$ H/8 $\times$ W/8   \\ 	\cline{2-7}				
				& Conv4    & 3 $\times$ 3       & 256      & 1          & 128 $\times$ H/8 $\times$ W/8  & 256 $\times$ H/8 $\times$ W/8  \\ \cline{2-7} 
				& MaxPooling     & 2 $\times$ 2       & -      & -          & 256 $\times$ H/8 $\times$ W/8 & 256 $\times$ H/16 $\times$ W/16  \\ \cline{2-7} 
				& ReLU     & -       & -       & -          & 256 $\times$ H/16 $\times$ W/16 & 256 $\times$ H/16 $\times$ W/16   \\ \cline{2-7}
				& Conv5     & 3 $\times$ 3        & 512      & 1          & 256 $\times$ H/16 $\times$ W/16  & 512 $\times$ H/16 $\times$ W/16  \\ \cline{2-7} 
				& MaxPooling     & 2 $\times$ 2       & -      & -          & 512 $\times$ H/16 $\times$ W/16 & 512 $\times$ H/32 $\times$ W/32  \\ \cline{2-7} 
				& ReLU     & -       & -       & -          & 512 $\times$ H/32 $\times$ W/32& 512 $\times$ H/32 $\times$ W/32   \\ \hline						
				
				\multirow{15}{*}{Decoder}           & Conv1     & 3 $\times$ 3       & 512      & 1          & 512 $\times$ H/32 $\times$ W/32& 512 $\times$ H/32 $\times$ W/32   \\ \cline{2-7} 
				& Upsampling     & 2 $\times$ 2       & -      & -          & 512 $\times$ H/32 $\times$ W/32  & 512 $\times$ H/16 $\times$ W/16  \\ \cline{2-7} 
				& ReLU     & -      & -       & -          & 512 $\times$ H/16 $\times$ W/16 & 512 $\times$ H/16 $\times$ W/16  \\ \cline{2-7}
				& Conv2    & 3 $\times$ 3       & 256      & 1          & 512 $\times$ H/16 $\times$ W/16 & 256 $\times$ H/16 $\times$ W/16  \\ \cline{2-7} 
				& Upsampling     & 2 $\times$ 2       & -      & -          & 256 $\times$ H/16 $\times$ W/16 & 256 $\times$ H/8 $\times$ W/8  \\ \cline{2-7} 
				& ReLU     & -       & -       & -          & 256 $\times$ H/8 $\times$ W/8 & 256 $\times$ H/8 $\times$ W/8   \\ \cline{2-7}
				& Conv3     & 3 $\times$ 3        & 128      & 1          & 256 $\times$ H/8 $\times$ W/8 & 128 $\times$ H/8 $\times$ W/8  \\ \cline{2-7} 
				& Upsampling     & 2 $\times$ 2       & -      & -          &  128 $\times$ H/8 $\times$ W/8 &  128 $\times$ H/4 $\times$ W/4  \\ \cline{2-7} 
				& ReLU     & -       & -       & -          & 128 $\times$ H/4 $\times$ W/4 & 128 $\times$ H/4 $\times$ W/4   \\ 	\cline{2-7}				
				& Conv2    & 3 $\times$ 3       & 64      & 1          & 128 $\times$ H/4 $\times$ W/4  & 64 $\times$ H/4 $\times$ W/4W  \\ \cline{2-7} 
				& Upsampling     & 2 $\times$ 2       & -      & -          & 64 $\times$ H/4 $\times$ W/4& 64 $\times$ H/2 $\times$ W/2  \\ \cline{2-7} 
				& ReLU     & -       & -       & -          & 64 $\times$ H/2 $\times$ W/2  & 64 $\times$ H/2 $\times$ W/2   \\ \cline{2-7}
				& Conv1     & 3 $\times$ 3        & 32      & 1          & 64 $\times$ H/2 $\times$ W/2  & 32 $\times$ H/2 $\times$ W/2 \\ \cline{2-7} 
				& Upsampling     & 2 $\times$ 2       & -      & -          & 32 $\times$ H/2 $\times$ W/2 & 32 $\times$ H $\times$ W \\ \cline{2-7} 
				& ReLU     & -       & -       & -          & 32 $\times$ H $\times$ W & 32 $\times$ H $\times$ W   \\ \hline	
			\end{tabular}
		}
	\end{center}
\end{table}

\begin{table}[h!]
	\caption{Run time comparison with SOTA methods}
	\label{Table:run_time}
	\centering
	\resizebox{1\linewidth}{!}{
		\begin{tabular}{|l|l|l|l|l|l|l|l|l|l|}
			\hline
			Method  & DSC  & LP   & JORDER & DDN  & JBO  & DID-MDN & SIRR & SPANet & Ours \\ \hline
			RunTime & 190s & 675s & 600s   & 0.3s & 1.4s & 0.2s    & 0.4s &0.3s&0.12s \\ \hline
	\end{tabular}}
	\vskip -8pt
\end{table} 

\bibliography{egbib}

\begin{thebibliography}{10}

\bibitem{yang2019single}
W.~Yang, R.~T. Tan, S.~Wang, Y.~Fang, and J.~Liu, ``Single image deraining:
  From model-based to data-driven and beyond,'' 2019.

\bibitem{zhang2019image}
H.~Zhang, V.~Sindagi, and V.~M. Patel, ``Image de-raining using a conditional
  generative adversarial network,'' {\em IEEE transactions on circuits and
  systems for video technology}, 2019.

\bibitem{krizhevsky2012imagenet}
A.~Krizhevsky, I.~Sutskever, and G.~E. Hinton, ``Imagenet classification with
  deep convolutional neural networks,'' in {\em Advances in neural information
  processing systems}, pp.~1097--1105, 2012.

\bibitem{szegedy2015going}
C.~Szegedy, W.~Liu, Y.~Jia, P.~Sermanet, S.~Reed, D.~Anguelov, D.~Erhan,
  V.~Vanhoucke, and A.~Rabinovich, ``Going deeper with convolutions,'' in {\em
  Proceedings of the IEEE conference on computer vision and pattern
  recognition}, pp.~1--9, 2015.

\bibitem{he2016deep}
K.~He, X.~Zhang, S.~Ren, and J.~Sun, ``Deep residual learning for image
  recognition,'' in {\em Proceedings of the IEEE conference on computer vision
  and pattern recognition}, pp.~770--778, 2016.

\bibitem{badrinarayanan2017segnet}
V.~Badrinarayanan, A.~Kendall, and R.~Cipolla, ``Segnet: A deep convolutional
  encoder-decoder architecture for image segmentation,'' {\em IEEE transactions
  on pattern analysis and machine intelligence}, vol.~39, no.~12,
  pp.~2481--2495, 2017.

\bibitem{ronneberger2015u}
O.~Ronneberger, P.~Fischer, and T.~Brox, ``U-net: Convolutional networks for
  biomedical image segmentation,'' in {\em International Conference on Medical
  image computing and computer-assisted intervention}, pp.~234--241, Springer,
  2015.

\bibitem{isola2017image}
P.~Isola, J.-Y. Zhu, T.~Zhou, and A.~A. Efros, ``Image-to-image translation
  with conditional adversarial networks,'' in {\em Proceedings of the IEEE
  conference on computer vision and pattern recognition}, pp.~1125--1134, 2017.

\bibitem{yi2017dualgan}
Z.~Yi, H.~Zhang, P.~Tan, and M.~Gong, ``Dualgan: Unsupervised dual learning for
  image-to-image translation,'' in {\em Proceedings of the IEEE international
  conference on computer vision}, pp.~2849--2857, 2017.

\bibitem{zhang2017style}
L.~Zhang, Y.~Ji, X.~Lin, and C.~Liu, ``Style transfer for anime sketches with
  enhanced residual u-net and auxiliary classifier gan,'' in {\em 2017 4th IAPR
  Asian Conference on Pattern Recognition (ACPR)}, pp.~506--511, IEEE, 2017.

\bibitem{wang2018high}
T.-C. Wang, M.-Y. Liu, J.-Y. Zhu, A.~Tao, J.~Kautz, and B.~Catanzaro,
  ``High-resolution image synthesis and semantic manipulation with conditional
  gans,'' in {\em Proceedings of the IEEE conference on computer vision and
  pattern recognition}, pp.~8798--8807, 2018.

\bibitem{iglovikov2018ternausnet}
V.~Iglovikov and A.~Shvets, ``Ternausnet: U-net with vgg11 encoder pre-trained
  on imagenet for image segmentation,'' {\em arXiv preprint arXiv:1801.05746},
  2018.

\bibitem{jin2017deep}
K.~H. Jin, M.~T. McCann, E.~Froustey, and M.~Unser, ``Deep convolutional neural
  network for inverse problems in imaging,'' {\em IEEE Transactions on Image
  Processing}, vol.~26, no.~9, pp.~4509--4522, 2017.

\bibitem{milletari2016v}
F.~Milletari, N.~Navab, and S.-A. Ahmadi, ``V-net: Fully convolutional neural
  networks for volumetric medical image segmentation,'' in {\em 2016 Fourth
  International Conference on 3D Vision (3DV)}, pp.~565--571, IEEE, 2016.

\bibitem{zhou2018unet++}
Z.~Zhou, M.~M.~R. Siddiquee, N.~Tajbakhsh, and J.~Liang, ``Unet++: A nested
  u-net architecture for medical image segmentation,'' in {\em Deep Learning in
  Medical Image Analysis and Multimodal Learning for Clinical Decision
  Support}, pp.~3--11, Springer, 2018.

\bibitem{cciccek20163d}
{\"O}.~{\c{C}}i{\c{c}}ek, A.~Abdulkadir, S.~S. Lienkamp, T.~Brox, and
  O.~Ronneberger, ``3d u-net: learning dense volumetric segmentation from
  sparse annotation,'' in {\em International conference on medical image
  computing and computer-assisted intervention}, pp.~424--432, Springer, 2016.

\bibitem{yasarla2020deblurring}
R.~Yasarla, F.~Perazzi, and V.~M. Patel, ``Deblurring face images using
  uncertainty guided multi-stream semantic networks,'' {\em IEEE Transactions
  on Image Processing}, vol.~29, pp.~6251--6263, 2020.

\bibitem{yasarla2020learning}
R.~Yasarla and V.~M. Patel, ``Learning to restore a single face image degraded
  by atmospheric turbulence using cnns,'' {\em arXiv preprint
  arXiv:2007.08404}, 2020.

\bibitem{fu2017removing}
X.~Fu, J.~Huang, D.~Zeng, Y.~Huang, X.~Ding, and J.~Paisley, ``Removing rain
  from single images via a deep detail network,'' in {\em Proceedings of the
  IEEE Conference on Computer Vision and Pattern Recognition}, pp.~3855--3863,
  2017.

\bibitem{zhang2018density}
H.~Zhang and V.~M. Patel, ``Density-aware single image de-raining using a
  multi-stream dense network,'' in {\em Proceedings of the IEEE conference on
  computer vision and pattern recognition}, pp.~695--704, 2018.

\bibitem{li2018recurrent}
X.~Li, J.~Wu, Z.~Lin, H.~Liu, and H.~Zha, ``Recurrent squeeze-and-excitation
  context aggregation net for single image deraining,'' in {\em Proceedings of
  the European Conference on Computer Vision (ECCV)}, pp.~254--269, 2018.

\bibitem{Wang_2019_CVPR}
T.~Wang, X.~Yang, K.~Xu, S.~Chen, Q.~Zhang, and R.~W. Lau, ``Spatial attentive
  single-image deraining with a high quality real rain dataset,'' in {\em The
  IEEE Conference on Computer Vision and Pattern Recognition (CVPR)}, June
  2019.

\bibitem{yasarla2019uncertainty}
R.~Yasarla and V.~M. Patel, ``Uncertainty guided multi-scale residual
  learning-using a cycle spinning cnn for single image de-raining,'' in {\em
  Proceedings of the IEEE Conference on Computer Vision and Pattern
  Recognition}, pp.~8405--8414, 2019.

\bibitem{lewicki2000learning}
M.~S. Lewicki and T.~J. Sejnowski, ``Learning overcomplete representations,''
  {\em Neural computation}, vol.~12, no.~2, pp.~337--365, 2000.

\bibitem{vincent2008extracting}
P.~Vincent, H.~Larochelle, Y.~Bengio, and P.-A. Manzagol, ``Extracting and
  composing robust features with denoising autoencoders,'' in {\em Proceedings
  of the 25th international conference on Machine learning}, pp.~1096--1103,
  2008.

\bibitem{kang2011automatic}
L.-W. Kang, C.-W. Lin, and Y.-H. Fu, ``Automatic single-image-based rain
  streaks removal via image decomposition,'' {\em IEEE transactions on image
  processing}, vol.~21, no.~4, pp.~1742--1755, 2011.

\bibitem{chen2014visual}
D.-Y. Chen, C.-C. Chen, and L.-W. Kang, ``Visual depth guided color image rain
  streaks removal using sparse coding,'' {\em IEEE transactions on circuits and
  systems for video technology}, vol.~24, no.~8, pp.~1430--1455, 2014.

\bibitem{luo2015removing}
Y.~Luo, Y.~Xu, and H.~Ji, ``Removing rain from a single image via
  discriminative sparse coding,'' in {\em Proceedings of the IEEE International
  Conference on Computer Vision}, pp.~3397--3405, 2015.

\bibitem{li2016rain}
Y.~Li, R.~T. Tan, X.~Guo, J.~Lu, and M.~S. Brown, ``Rain streak removal using
  layer priors,'' in {\em Proceedings of the IEEE conference on computer vision
  and pattern recognition}, pp.~2736--2744, 2016.

\bibitem{zhang2017convolutional}
H.~Zhang and V.~M. Patel, ``Convolutional sparse and low-rank coding-based rain
  streak removal,'' in {\em 2017 IEEE Winter Conference on Applications of
  Computer Vision (WACV)}, pp.~1259--1267, IEEE, 2017.

\bibitem{wang2017hierarchical}
Y.~Wang, S.~Liu, C.~Chen, and B.~Zeng, ``A hierarchical approach for rain or
  snow removing in a single color image,'' {\em IEEE Transactions on Image
  Processing}, vol.~26, no.~8, pp.~3936--3950, 2017.

\bibitem{zhang2006rain}
X.~Zhang, H.~Li, Y.~Qi, W.~K. Leow, and T.~K. Ng, ``Rain removal in video by
  combining temporal and chromatic properties,'' in {\em 2006 IEEE
  International Conference on Multimedia and Expo}, pp.~461--464, IEEE, 2006.

\bibitem{garg2007vision}
K.~Garg and S.~K. Nayar, ``Vision and rain,'' {\em International Journal of
  Computer Vision}, vol.~75, no.~1, pp.~3--27, 2007.

\bibitem{santhaseelan2015utilizing}
V.~Santhaseelan and V.~K. Asari, ``Utilizing local phase information to remove
  rain from video,'' {\em International Journal of Computer Vision}, vol.~112,
  no.~1, pp.~71--89, 2015.

\bibitem{liu2018erase}
J.~Liu, W.~Yang, S.~Yang, and Z.~Guo, ``Erase or fill? deep joint recurrent
  rain removal and reconstruction in videos,'' in {\em Proceedings of the IEEE
  Conference on Computer Vision and Pattern Recognition}, pp.~3233--3242, 2018.

\bibitem{li2018video}
M.~Li, Q.~Xie, Q.~Zhao, W.~Wei, S.~Gu, J.~Tao, and D.~Meng, ``Video rain streak
  removal by multiscale convolutional sparse coding,'' in {\em Proceedings of
  the IEEE Conference on Computer Vision and Pattern Recognition},
  pp.~6644--6653, 2018.

\bibitem{yang2019frame}
W.~Yang, J.~Liu, and J.~Feng, ``Frame-consistent recurrent video deraining with
  dual-level flow,'' in {\em Proceedings of the IEEE Conference on Computer
  Vision and Pattern Recognition}, pp.~1661--1670, 2019.

\bibitem{ren2019progressive}
D.~Ren, W.~Zuo, Q.~Hu, P.~Zhu, and D.~Meng, ``Progressive image deraining
  networks: a better and simpler baseline,'' in {\em Proceedings of the IEEE
  Conference on Computer Vision and Pattern Recognition}, pp.~3937--3946, 2019.

\bibitem{wang2019erl}
G.~Wang, C.~Sun, and A.~Sowmya, ``Erl-net: Entangled representation learning
  for single image de-raining,'' in {\em Proceedings of the IEEE International
  Conference on Computer Vision}, pp.~5644--5652, 2019.

\bibitem{deng2020detail}
S.~Deng, M.~Wei, J.~Wang, Y.~Feng, L.~Liang, H.~Xie, F.~L. Wang, and M.~Wang,
  ``Detail-recovery image deraining via context aggregation networks,'' in {\em
  Proceedings of the IEEE/CVF Conference on Computer Vision and Pattern
  Recognition}, pp.~14560--14569, 2020.

\bibitem{wang2020model}
H.~Wang, Q.~Xie, Q.~Zhao, and D.~Meng, ``A model-driven deep neural network for
  single image rain removal,'' in {\em Proceedings of the IEEE/CVF Conference
  on Computer Vision and Pattern Recognition}, pp.~3103--3112, 2020.

\bibitem{jiang2020multi}
K.~Jiang, Z.~Wang, P.~Yi, C.~Chen, B.~Huang, Y.~Luo, J.~Ma, and J.~Jiang,
  ``Multi-scale progressive fusion network for single image deraining,'' in
  {\em Proceedings of the IEEE/CVF Conference on Computer Vision and Pattern
  Recognition}, pp.~8346--8355, 2020.

\bibitem{yasarla2020confidence}
R.~Yasarla and V.~M. Patel, ``Confidence measure guided single image
  de-raining,'' {\em IEEE Transactions on Image Processing}, vol.~29,
  pp.~4544--4555, 2020.

\bibitem{du2020conditional}
Y.~Du, J.~Xu, X.~Zhen, M.-M. Cheng, and L.~Shao, ``Conditional variational
  image deraining,'' {\em IEEE Transactions on Image Processing}, 2020.

\bibitem{wang2020rethinking}
Y.~Wang, Y.~Song, C.~Ma, and B.~Zeng, ``Rethinking image deraining via rain
  streaks and vapors,'' {\em arXiv preprint arXiv:2008.00823}, 2020.

\bibitem{liu2012robust}
G.~Liu, Z.~Lin, S.~Yan, J.~Sun, Y.~Yu, and Y.~Ma, ``Robust recovery of subspace
  structures by low-rank representation,'' {\em IEEE transactions on pattern
  analysis and machine intelligence}, vol.~35, no.~1, pp.~171--184, 2012.

\bibitem{mairal2009online}
J.~Mairal, F.~Bach, J.~Ponce, and G.~Sapiro, ``Online dictionary learning for
  sparse coding,'' in {\em Proceedings of the 26th annual international
  conference on machine learning}, pp.~689--696, 2009.

\bibitem{fu2017clearing}
X.~Fu, J.~Huang, X.~Ding, Y.~Liao, and J.~Paisley, ``Clearing the skies: A deep
  network architecture for single-image rain removal,'' {\em IEEE Transactions
  on Image Processing}, vol.~26, no.~6, pp.~2944--2956, 2017.

\bibitem{hu2019depth}
X.~Hu, C.-W. Fu, L.~Zhu, and P.-A. Heng, ``Depth-attentional features for
  single-image rain removal,'' in {\em Proceedings of the IEEE Conference on
  Computer Vision and Pattern Recognition}, pp.~8022--8031, 2019.

\bibitem{halder2019physics}
S.~S. Halder, J.-F. Lalonde, and R.~d. Charette, ``Physics-based rendering for
  improving robustness to rain,'' in {\em Proceedings of the IEEE International
  Conference on Computer Vision}, pp.~10203--10212, 2019.

\bibitem{wei2019semi}
W.~Wei, D.~Meng, Q.~Zhao, Z.~Xu, and Y.~Wu, ``Semi-supervised transfer learning
  for image rain removal,'' in {\em Proceedings of the IEEE Conference on
  Computer Vision and Pattern Recognition}, pp.~3877--3886, 2019.

\bibitem{yasarla2020syn2real}
R.~Yasarla, V.~A. Sindagi, and V.~M. Patel, ``Syn2real transfer learning for
  image deraining using gaussian processes,'' in {\em Proceedings of the
  IEEE/CVF Conference on Computer Vision and Pattern Recognition},
  pp.~2726--2736, 2020.

\bibitem{8767931}
X.~{Fu}, B.~{Liang}, Y.~{Huang}, X.~{Ding}, and J.~{Paisley}, ``Lightweight
  pyramid networks for image deraining,'' {\em IEEE Transactions on Neural
  Networks and Learning Systems}, vol.~31, no.~6, pp.~1794--1807, 2020.

\bibitem{WangA}
H.~Wang, Y.~Wu, M.~Li, Q.~Zhao, and D.~Meng, ``A survey on rain removal from
  video and single image,'' {\em arXiv preprint arXiv:1909.08326}, 2019.

\bibitem{8627954}
W.~{Yang}, R.~T. {Tan}, J.~{Feng}, Z.~{Guo}, S.~{Yan}, and J.~{Liu}, ``Joint
  rain detection and removal from a single image with contextualized deep
  networks,'' {\em IEEE Transactions on Pattern Analysis and Machine
  Intelligence}, vol.~42, no.~6, pp.~1377--1393, 2020.

\bibitem{nair2010rectified}
V.~Nair and G.~E. Hinton, ``Rectified linear units improve restricted boltzmann
  machines,'' in {\em Proceedings of the 27th international conference on
  machine learning (ICML-10)}, pp.~807--814, 2010.

\bibitem{johnson2016perceptual}
J.~Johnson, A.~Alahi, and L.~Fei-Fei, ``Perceptual losses for real-time style
  transfer and super-resolution,'' in {\em European conference on computer
  vision}, pp.~694--711, Springer, 2016.

\bibitem{yang2017deep}
W.~Yang, R.~T. Tan, J.~Feng, J.~Liu, Z.~Guo, and S.~Yan, ``Deep joint rain
  detection and removal from a single image,'' in {\em Proceedings of the IEEE
  Conference on Computer Vision and Pattern Recognition}, pp.~1357--1366, 2017.

\bibitem{kingma2014adam}
D.~P. Kingma and J.~Ba, ``Adam: A method for stochastic optimization,'' {\em
  arXiv preprint arXiv:1412.6980}, 2014.

\end{thebibliography}
\bibliographystyle{ieeetr}

\end{document}